\documentstyle[mncite,psfig]{mn}

\newcommand{\ltae}{\raisebox{-0.6ex}{$\,\stackrel
{\raisebox{-.2ex}{$\textstyle <$}}{\sim}\,$}}

\title[Jet-cloud interactions in high-z radio galaxies]
{The evidence for jet-cloud interactions in a sample of high/intermediate-redshift radio galaxies}

\author[C.\ Sol\'{o}rzano-I\~{n}arrea et\ al.]
{C. Sol\'{o}rzano-I\~{n}arrea$^{1,2}$, C. N. Tadhunter$^1$,
D. J. Axon$^3$\\
$^{1}$Department of Physics and Astronomy, University of Sheffield,
Sheffield, S3 7RH\\
$^{2}$Department of Physics and Astronomy, University of Leeds, Leeds,
LS2 9JT\\
$^{3}$Department of Physical Sciences, University of Hertfordshire, Hatfield, AL10 9AB\\}
\begin{document}
\maketitle
\begin{abstract}
We present detailed long-slit spectroscopic observations of a sample of four powerful radio galaxies, with intermediate/high redshifts (0.47 $<$ z $<$ 0.81). The observations cover the rest-wavelength range 2600 \AA - 6600 \AA, chosen to include important diagnostic emission lines ([NeV]$\lambda3426$,
[OII]$\lambda3727$, [NeIII]$\lambda3869$, H$\beta$, [OIII]$\lambda5007$), which are also measured in optical observations of low-redshift radio galaxies. In two of the galaxies (3C 352 and 3C 435A) the radio sources are of the same scale as the emission-line regions, whereas in the other two (3C 34 and 3C 330) the radio sources are extended on a larger scale than the emission-line structures. We find that the extended regions of \emph {all} the galaxies present highly disturbed kinematics, consisting of line-splitting ($\Delta v \sim$ 1000 km s$^{-1}$) and/or underlying broad components (FWHM = 1000 --- 1500 km s$^{-1}$). These features are difficult to explain in terms of gravitational motion in the haloes of the host galaxies. Rather, it is likely that they are the result of strong shocks driven through the ISM/IGM by the radio sources. These observations suggest that jet-induced shocks have an important effect on the emission-line properties even in sources in which the radio structures are on a much larger scale than the emission-line structures. 

While the emission-line kinematics provide strong evidence for shock acceleration, the dominant ionization mechanism for the emission-line gas remains 
uncertain. We have compared the optical diagnostic line ratios of the galaxies in our sample with various ionization models, including: pure shock ionization, shocks including a photoionized precursor, power-law photoionization and photoionization including matter-bounded clouds. We find that both pure-shock ionization and power-law photoionization model predictions fail to provide good fits to the data. On the other hand, on individual diagnostic diagrams, models for shocks which include a photoionized precursor are consistent with the results for the majority of the EELR of the galaxies in our sample, and photoionization including matter-bounded clouds models also give reasonable fits to some of the EELR. However, in terms of the positions of the points relative to the model sequences on the diagnostic diagrams, there is a lack of consistency from diagram to diagram. The diagnostic diagram involving the line ratios [OIII](4959+5007)/4363 and HeII(4686)/H$\beta$ is particularly problematic in this regard. Overall, our results suggest that, if the EELR are shock-ionized, one or more of the assumptions implicit in the shock models may need to be reconsidered.

In addition, we have investigated the nebular continuum contribution to the UV excess in the galaxies in our sample. We find a substantial nebular emission contribution to the UV continuum in all the cases, in the range $\sim$ 10 --- 40 per cent. However, after the subtraction of the nebular component, a significant UV excess remains in the extended nebulae of most of the objects.   

\end{abstract}

\newpage
\section{Introduction}
Most powerful radio galaxies have extended emission-line regions (EELR) extending to large distances from their nuclei (5 --- 100 kpc) \cite{tadhunter86}. These EELR offer the possibility of investigating the origin of the activity and the origin of the gas itself. However, if we are going to use the EELR in this way, we need to know the extent to which the emission-line morphologies and kinematics reflect the intrinsic properties of the gaseous haloes, and the extend to which they reflect the effects of the activity. Thus, it is crucial to understand the ionization mechanisms in these sources.
 
The nature of the mechanisms involved in producing the observed kinematics and ionization of the EELR remains highly uncertain. Currently, the two most accepted models are photoionization by a central active nucleus, and shock-ionization by the interaction between the radio-emitting components and the ambient gas. In the latter model, the so-called jet-cloud interaction model, the advancing radio jet drives strong shocks into the warm clouds of gas in the haloes of the host galaxies. The regions directly behind the shocks are compressed and heated to very high temperatures and line emission is produced by two distinct mechanisms: (a) as the post-shock clouds cool by emitting line radiation, and (b) photoionization of the cool, dense, post-shock gas and the precursor gas ahead of the shock by the hot shocked gas. It is likely that no single mechanism, but a combination of both AGN photoionization and jet-cloud interaction is required to explain the kinematics and ionization of the EELR over the whole range of redshifts.

At low redshifts (z $<$ 0.3), the properties of the EELR in most radio galaxies appear to be consistent with the illumination of the ambient gas by the radiation field of an active nucleus (e.g. Fosbury 1989), and the emission-line kinematics can be explained in terms of pure gravitational motion in the haloes of the galaxies (\pcite{tadhunter89b}; \pcite{baum90}). This illumination model is in agreement with the unified schemes (e.g. \pcite{barthel89}), in which it is proposed that all active galaxies are intrinsically identical but, due to orientation effects, their appearance varies.

However, some nearby radio galaxies show clear evidence for strong interactions between the radio-emitting components and the ambient gas. Features observed in these galaxies include: 
\begin{itemize}
 \item  close morphological association between radio and optical emission;
 \item  complex kinematics (underlying broad components and split narrow components);
 \item  large linewidths in the extended gas;
 \item  anticorrelations between linewidth and ionization state.
\end{itemize}

Detailed studies by Clark et al.(1997, 1998) and \scite{villarmartin99b} present direct evidence for the dramatic effect that shocks can have in the properties of the extended emission-line gas in a sample of low-redshift radio galaxies. The emission lines of these galaxies are resolved in two main kinematic components: a low-ionization broad component, spatially associated with the radio structures, and a high-ionization narrow component, extending beyond the radio structures. Furthermore, a recent study of two nearby radio galaxies \cite{tadhunter2000} has revealed low-surface-brightness emission-line structures at large distances from the radio jet axis. These structures are inconsistent with the illumination model, since parts of them lie outside the ionization cones predicted by the unified schemes, and rather suggest that the ionizing effects of the radio components can extend far from the radio jet axes. 

At higher redshifts (z $>$ 0.5) there appears to be more evidence for shocks than in the population of low-redshift radio galaxies. At such redshifts powerful radio galaxies frequently show highly collimated UV continuum and emission-line structures which are closely aligned with the radio axes (\pcite{mccarthy87}; \pcite{chambers87}; \pcite{best96}). Such collimated structures are not what would be expected on the basis of illumination by the broad radiation cones predicted by the unified schemes. In addition, many high-redshift radio galaxies present highly disturbed emission-line kinematics along the radio axis, which are difficult to explain in terms of pure gravitational motions \cite{mccarthy96}. This suggests that processes other than AGN illumination may be at work. Indeed, many of these features observed in powerful high-redshift radio galaxies are consistent with the jet-cloud interaction model described above and with the results of detailed studies of jet-cloud interactions in low-redshift radio galaxies.

In addition, recent work by \scite{best2000a}, at redshifts z $\simeq$ 1, has demonstrated that there is a strong evolution of the emission-line properties of powerful radio galaxies with radio size. They find that radio sources with small linear sizes have emission-line properties which are in agreement with the theoretical shock ionization predictions, and show highly distorted velocity profiles. In contrast, the larger radio sources, where the shock fronts have passed beyond the emission-line regions, present emission-line ratios consistent with AGN photoionization, with higher ionization states and narrower line widths than those of the smaller radio sources.

However, in apparent contradiction with the results obtained at z $\simeq$ 1, \scite{villarmartin97}, basing their study on the UV line ratios of a sample of radio galaxies at z$>$2, found that the EELR appear to be mainly photoionized by the central AGN in the very high redshift regime.
 
One problem with comparing the results from samples at different redshifts is that optical observations sample different rest wavelengths as the redshift varies. Consequently, the diagnostic diagrams used to determine the ionization mechanism involve lines of different ionization and excitation as the redshift changes, and there is a danger that our sensitivity to particular ionization mechanisms varies with wavelength. Because of this problem, it has been difficult to determine the balance between shocks and photoioinization, and how this balance varies with redshift.
 
In this paper we present detailed long-slit spectroscopic observations of four powerful high-redshift (0.47 $<$ z $<$ 0.81) radio galaxies, aimed at investigating the balance between shocks and AGN-photoionization. With these observations we extend the approach of combining kinematical and ionization information --- previously used in studies of low-redshift radio galaxies (Clark et al. 1997, 1998; \pcite{villarmartin99b}) --- to the higher redshift regimes. Crucially, these new data cover the same rest wavelength range as previous studies of low-redshift radio galaxies, thereby allowing direct comparisons to be made between low and high-redshift objects.
\section{Sample selection}
Our sample consists of four 3C radio galaxies imaged by \scite{mccarthy95}, with intermediate redshifts in the range 0.47 $<$ z $<$ 0.81. The upper redshift limit was chosen to ensure that the rest-wavelength range 3400 -- 5200 \AA \ was accessible with optical instrumentation. The galaxies are the following: 3C 34 (z=0.690), 3C 330 (z=0.550), 3C 352 (z=0.8067) and 3C 435A (z=0.471). All four objects were previously known as having extended emission-line regions aligned along their radio axes. We also selected the sources so as to have different radio sizes relative to the corresponding host galaxies. 3C 34 and 3C 330 are large double radio sources, in which the radio hot spots have passed far beyond the EELR. 3C 352 and 3C 435A are medium-size double radio sources with close associations between the radio and optical structures, and with emission line and radio emission on a similar scale. Although 3C 352 is the highest redshifted galaxy of our sample, its EELR are large enough to allow a spatially resolved analysis. 
\section{Observations, data reduction and analysis}
Long-slit spectroscopic observations were carried out on the night 15/8/93 for 3C 34 and on the nights 20-21-22/7/96 for the rest of the sample, using the ISIS double-armed spectrograph on the 4.2m William Herschel Telescope on La Palma (Spain). Details of these observations are presented in Table~\ref{logobs}. Blue and red spectra were taken using the Tek1 (0.357 arcsec/pix) and EEV3 (0.335 arcsec/pix) CCD detectors, respectively. The slit was oriented approximately parallel to the radio axis for each galaxy (see Table~\ref{logobs} for details).

The data-reduction was carried out using the IRAF software package and was performed in several stages: bias-subtraction (using the unilluminated region of the chip); cosmic-ray removal (manually using the {\footnotesize IMEDIT} task); flat-fielding (using the normalized flat-field frame); wavelength calibration (using copper-argon comparison frames for the blue spectra and copper-neon for the red spectra); atmospheric extinction correction, flux calibration and sky subtraction. From measurements of night-sky emission lines, the absolute uncertainties in the wavelength calibration are estimated to be $\sim$ 0.2 \AA \ for the 300B grating, and $\sim$ 0.4 \AA \ for the 158B and 158R gratings. Comparison between different spectrophotometric standard star observations gives a flux calibration error of $\sim$ 5 per cent across the wavelength range within each data set. Following the basic reduction, the two-dimensional spectra were corrected for any distortion along the spectral direction, using the Starlink FIGARO package. Then, the red spectra of all the galaxies, except 3C 34, were resampled in the spatial direction to have the same pixel scale as the blue spectra (0.357 arcsec/pix).  

The two-dimensional spectra were all shifted to the corresponding rest frame of each galaxy before the analysis. One-dimensional spectra were then extracted, and analysed using the Starlink DIPSO spectral analysis package. Gaussians were fitted to the profiles of the emission lines. The measured line widths were all corrected for the spectral resolution of the instrument via the Gaussian quadrature technique $(w=[w^{2}_{obs}-w_{inst}^{2}]^{\frac{1}{2}})$. The instrumental widths (corresponding to the instrumental resolution), derived using the night-sky lines and verified with arc lines, are listed in Table~\ref{logobs}.

The data have not been corrected for Galactic reddening. None of the objects in our sample have low latitude and, therefore, the local extinction is not large for any of them (0.022 $\leq$ E(B-V) $\leq$ 0.059).

A Hubble constant of H$_{0}$ = 50 km s$^{-1}$ Mpc$^{-1}$ and a deceleration parameter of q$_{0}$ = 0.5 are assumed throughout this paper. Details of the resulting angular scale for each galaxy are given in Table~\ref{angscale}.
\section{Results}
\subsection{3C 34}
\subsubsection{Previous observations}
3C 34 is a large 46 arcsec (365 kpc) classical FRII double radio source, at a redshift z = 0.69. It was first optically identified by \scite{riley80}. Imaging by \scite{mccarthy95} reveals that the galaxy lies in a rich compact cluster environment, and an extended [OII]$\lambda$3727 emission region of high-surface brightness aligned with the radio source. A recent analysis of the environment around 3C 34, based on K-band images, confirms that the galaxy resides in a rich cluster environment \cite{best2000b}. Despite the fact that the radio source is larger than the emission-line nebula, \scite{best97} report possible evidence for a strong jet-induced star formation. Their Hubble Space Telescope images of 3C 34 show a long narrow region of blue emission orientated along the radio axis and directed towards the radio hot spot. They propose that this aligned emission is associated with a region of massive star formation, induced by the passage of the radio jet through a galaxy within the cluster surrounding 3C 34. Based on Fabry-P\'{e}rot images, \scite{neeser97} propose that the extended emission-line region in 3C 34 is photoionized by anisotropic UV radiation emitted by the central AGN. They find that this interpretation is energetically viable, but insufficient to explain the line-emission kinematics observed in 3C 34. Therefore, they suggest that the kinematics in the extended regions of 3C 34 are best explained as gas swept aside by the lateral expansion of the radio source lobes.
\subsubsection{The emission-line structure}
\label{contsubtraction}
Since the airmass was low (1.073) for the observations of 3C 34 and the slit wide, the effects of differential atmospheric refraction can be neglected. The spatial profiles of [OII]$\lambda3727$ (solid line) and [OIII]$\lambda5007$ (dashed line) are presented in Figure~\ref{34linecont} (top). They were derived in the following way: a 40\AA \ wide slice, perpendicular to the dispersion direction and centred on the appropriate emission line, was extracted; then, to correct for the continuum contamination, a slice of the same width, taken from an adjacent line-free region of the spectrum, was subtracted from the first one. 

In these profiles, line emission is seen extending approximately 16 arcsec (127 kpc) along the radio axis. The positions of the radio hot spots lie far outside the EELR, and therefore they are not indicated on the plots. In the nucleus, the [OII] emission peak is coincident with the continuum centroid, while the [OIII] emission peaks slightly to the east. In addition to the central nuclear peak, the spatial profiles show a pronounced peak $\sim$ 4.5 arcsec west of the nucleus. It is noticeable that the [OII] emission is stronger than the [OIII] emission in both the nucleus and the western EELR. This indicates a low-ionization state.   
\subsubsection{The emission-line kinematics}
\begin{figure}
\psfig{figure=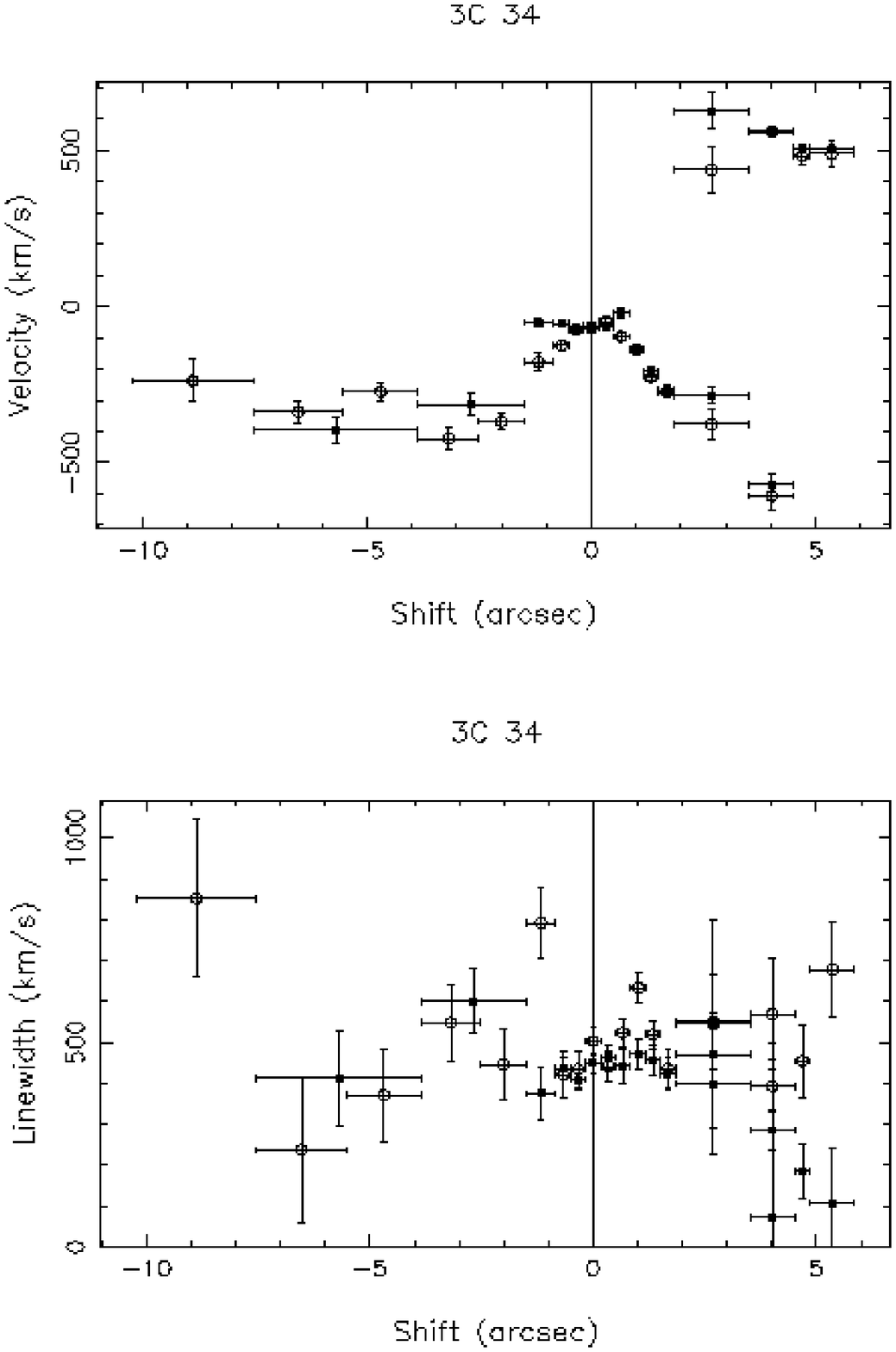,width=8.5cm}
\caption{Variation in the rest-frame radial velocity and in the linewidths (FWHM) along the radio axis of 3C 34 for [OII] (open circles) and [OIII] (filled 
squares).}
\label{velwidth34}
\end{figure}

\begin{figure}
\psfig{figure=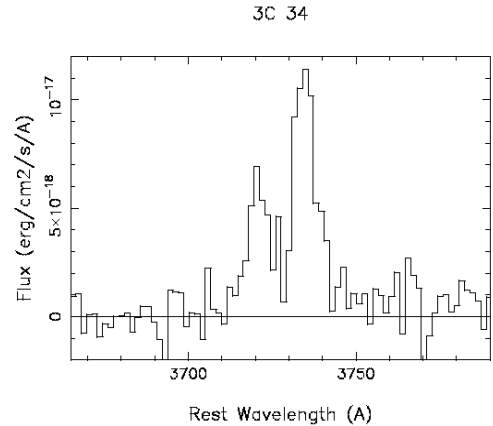,angle=-90,width=8.5cm}
\caption{Extracted spectrum of 3C 34 for 1.7 x 1.5 arcsec$^{2}$ aperture centred $\sim$ 3.7 arcsec west of the continuum centroid, showing the split components in the 
[OII] line.}
\label{split34}
\end{figure}

Kinematic information was extracted from the one-dimensional spectra by fitting single Gaussians to the profiles of the strong emission lines [OII]$\lambda3727$ and [OIII]$\lambda5007$. Figure~\ref{velwidth34} (top) shows the variation in the velocity centroids of these emission lines along the radio axis. Both the [OII] and [OIII] velocities vary in a similar manner, although the [OIII] line emission does not extend to the east as far as the [OII] line emission. A strong splitting ($\sim$ 1000 km~s$^{-1}$) in both [OII] and [OIII] lines is observed at a radial distance of 2   to 4.5 arcsec (16 to 36 kpc) on the west side of the nucleus along the radio axis. It can be seen that the splitting is almost symmetric with respect to the nuclear velocity, having the two components relative shifts of $\sim$ + 500 and $\sim$ -- 500 km s$^{-1}$. The extracted spectrum (1.7 x 1.5 arcsec$^{2}$ aperture centred $\sim$ 3.7 arcsec west of the continuum centroid) showing the high-velocity components to the [OII] line is presented in Figure~\ref{split34}.

The variations in the linewidths along the radio axis for the [OII] and [OIII] emission lines are shown in Figure~\ref{velwidth34} (bottom). The linewidths are found to be moderate, about 400 km~s$^{-1}$ on average, for both [OII] and [OIII] along the direction of the slit. Across much of the galaxy, the linewidths of [OII] and [OIII] are consistent, except for the small regions at $\sim$ 1 arcsec distance from the nucleus at both sides, where there are small but significant radial velocity differences between [OII] and [OIII]. Also, on the west side of the nucleus, at the location of the splitting, it can be seen that the [OII] line tends to be slightly broader than the [OIII] line; however, this could be due to the fact that the [OII] line is a doublet ([OII]$\lambda$3726.0 and [OII]$\lambda$3728.8).

These results agree with those obtained by \scite{neeser97}, based on Fabry-P\'{e}rot imaging. Although they do not detect the line splitting, the large linewidths ($\sim$ 1200 km s$^{-1}$) they find in the western EELR are consistent with unresolved splitting. They explain this as being the overlap of two distinct sources (the central one and the western extended region).
\subsubsection{The emission-line spectra}
\begin{figure*}
\begin{minipage}{160mm}
\psfig{figure=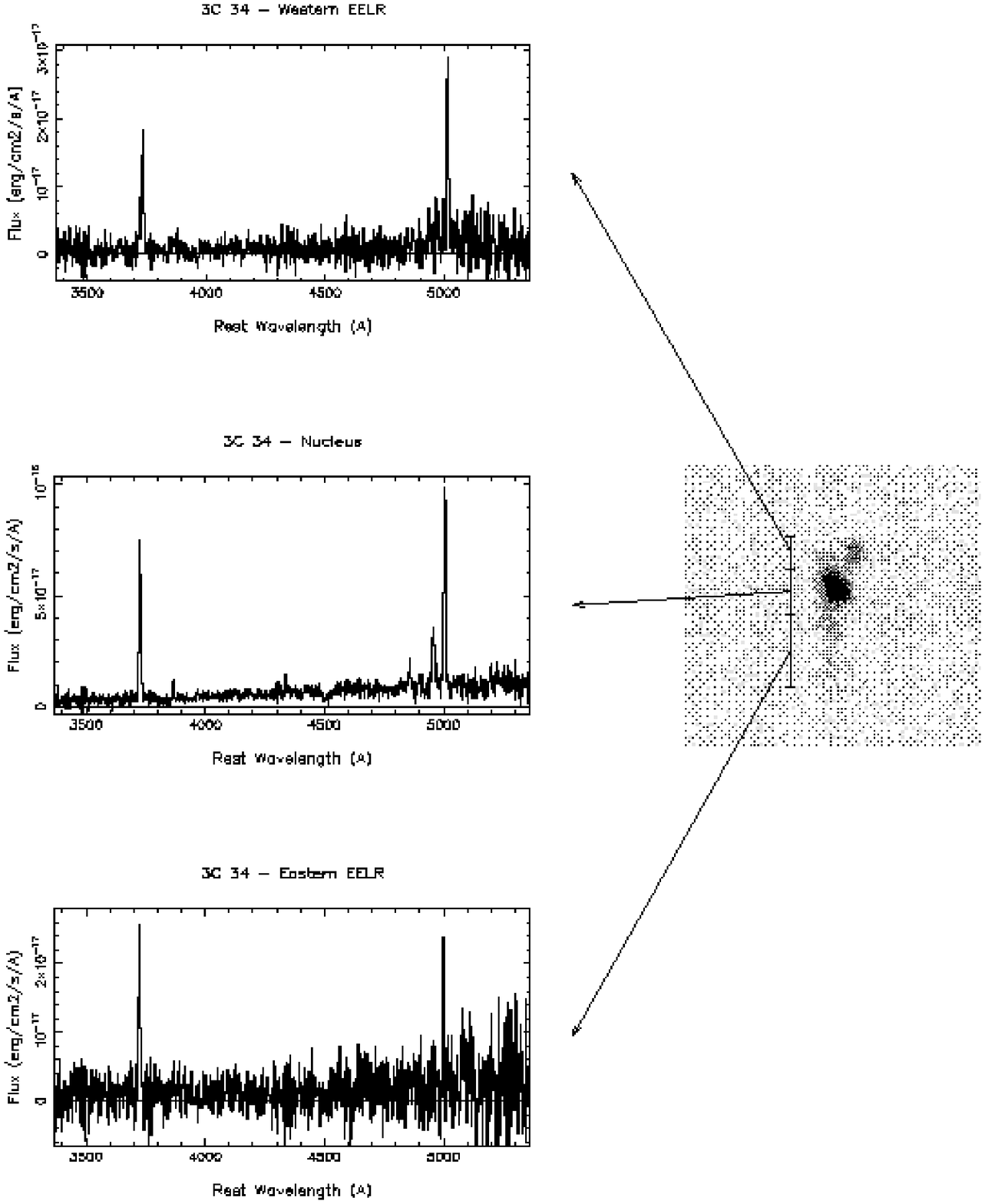,width=16cm}
\caption{Integrated blue and red spectra of the different regions in 3C 34, and 2-D spectrum showing [OII]$\lambda$3727 (3660 \AA \ -- 3790 \AA, 28.8 
arcsec along the slit).}
\label{34specim}
\end{minipage}
\end{figure*}

Integrated blue and red spectra of the eastern EELR, the central nuclear region and the western EELR of 3C 34, and a 2-D spectrum showing the [OII]$\lambda$3727 emission line are presented in Figure~\ref{34specim}. The integrated fluxes of the observed emission lines in these three regions are given in Table~\ref{tabflux34}. The fluxes are normalised to the observed-frame H$\beta$ flux for each aperture. The fluxes have not been corrected for intrinsic reddening, given that the H$\gamma$/H$\beta$ ratio is consistent with the theoretical value corresponding to Case B recombination \cite{osterbrock89}.
 
\begin{figure}
\psfig{figure=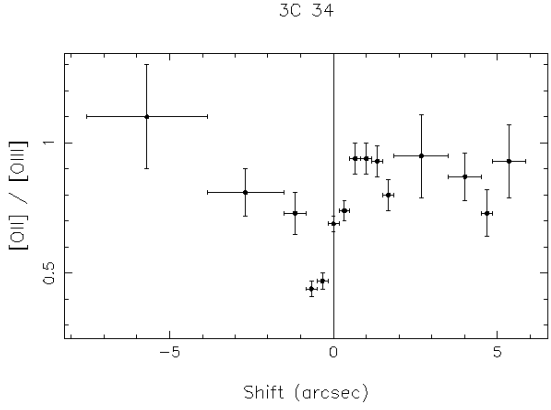,angle=-90,width=8.5cm}
\caption{Variation in the [OII]/[OIII] line ratio along the radio axis of 3C 34.}
\label{34ro2o3}
\end{figure}

Figure~\ref{34ro2o3} shows the variation in the [OII]($\lambda3727$) / [OIII]($\lambda5007$) line ratio along the radio axis. Apart from a high-ionization region to the east of the nucleus, the spectra show consistently a low ionization state over most of the nebula ([OII]/[OIII] $>$ 0.7).
\subsubsection{The physical conditions}
Unfortunately, it was not possible to measure the electron temperature for 3C 34 from the [OIII](4959+5007)/4363 because the [OIII]$\lambda4363$ emission line was not detected.
\subsubsection{The continuum emission}
\label{nebcont}
It has been demonstrated \cite{dickson95} that the nebular continuum emission (a combination of free-free, free-bound recombination and two-photon continua, plus higher order Balmer lines) makes a major contribution to the UV continua of powerful radio galaxies wherever the emission lines have large equivalent widths. However, there is still some controversy about the relative importance of the nebular component in the UV continuum emission. The theoretical Case B nebular continuum can be generated by using the H$\beta$ flux (e.g. \pcite{osterbrock89}). This calculation is not particularly sensitive to density and temperature variations, only the intrinsic reddening could give some uncertainty; but this is not a problem since no significant reddening has been measured in 3C 34.

The spatial profile of the nebular continuum was calculated in the following way: a spatial slice, perpendicular to the dispersion direction, containing the H$\beta$ emission line was extracted from the rest-frame 2-D spectrum. Then, an adjacent continuum slice of the same width was subtracted from the first one to obtain the continuum-free H$\beta$ profile, which was used to produce the nebular continuum profile (by using the {\small NEBCONT} routine in DIPSO).

\begin{figure}
\psfig{figure=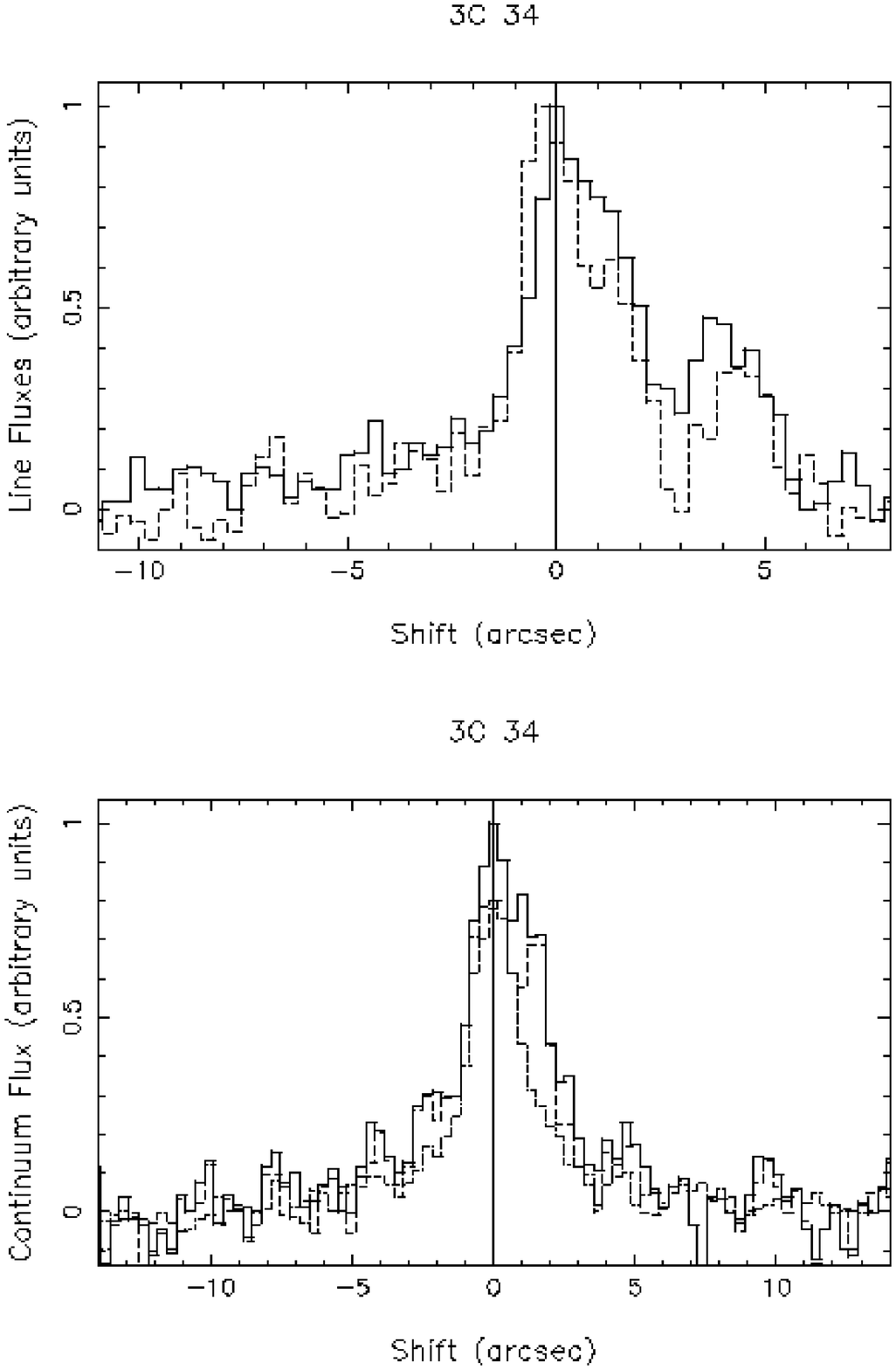,width=8.5cm}
\caption{Top: Variation of [OII] (solid line) and [OIII] (dashed line) fluxes along the radio axis of 3C 34. Bottom: Profile of the line-free continuum 
emission along the radio axis of 3C 34. The solid line shows the 3400 \AA \ -- 3700 \AA \ profile, the dashed line shows the blue-continuum emission after the 
subtraction of the nebular contribution (calculated from the H$\beta$ flux along the slit), and the dotted line shows the 4500\AA \ -- 4800 \AA \ profile 
(scaled to the peak of the nebular-subtracted blue profile). (West is to the right)}
\label{34linecont}
\end{figure}

Figure~\ref{34linecont} (bottom) presents the spatial profile of the continuum emission along the radio axis of 3C 34. The solid line shows the UV continuum emission from 3400 \AA \ -- 3700 \AA, the dashed line shows the blue-continuum emission after the subtraction of the nebular continuum contribution, and the dotted line shows the continuum emission from 4500 \AA \ -- 4800 \AA, scaled to the peak of the nebular-subtracted blue profile. It is found that the contribution of the nebular emission is $\sim$ 12 per cent on the nucleus and $\sim$ 11 per cent on the eastern EELR, for the wavelength range 3400 \AA \ -- 3700 \AA. From Figure~\ref{34linecont} (bottom), it can be seen that the nebular continuum cannot account for the UV continuum excess observed on both sides of the nucleus, especially on the west side.
\subsection{3C 330}
\subsubsection{Previous observations}
3C 330, with a redshift z = 0.550, was first optically identified by \scite{spinrad76}. It is a large 62 arcsec (458 kpc) double radio source. Ground-based [OIII]$\lambda$5007 and radio continuum images \cite{mccarthy91} reveal that 3C 330 has a small lobe distance asymmetry, yet shows a large asymmetry in its extended emission lines, which are stronger on the side of the radio lobe closer to the nucleus (North-East). More recent HST narrow-band images show a narrow cone-like structure on the north-east side of the galaxy, which has been interpreted in terms of illumination by a quasar hidden in the core of the galaxy \cite{mccarthy97}. However, the putative ``ionization cone'' is narrower than expected on the basis of the broad cones predicted by the unified schemes for powerful radio galaxies \cite{barthel89}. Continuum images by \scite{mccarthy95} indicate that this object lies in a cluster and has a close companion to the north. In contrast, ROSAT observations of 3C 330 reveal a relatively low X-ray luminosity \cite{crawford96}, which is not compatible with the presence of a rich cluster in the environment of 3C 330. Based on long-slit spectroscopy, \scite{mccarthy96} carried out kinematic analysis of 3C 330, finding a large, smooth velocity gradient along PA 230$^{o}$ in [OIII]$\lambda\lambda$4959,5007.
\subsubsection{The emission-line structure}
\begin{figure}
\psfig{figure=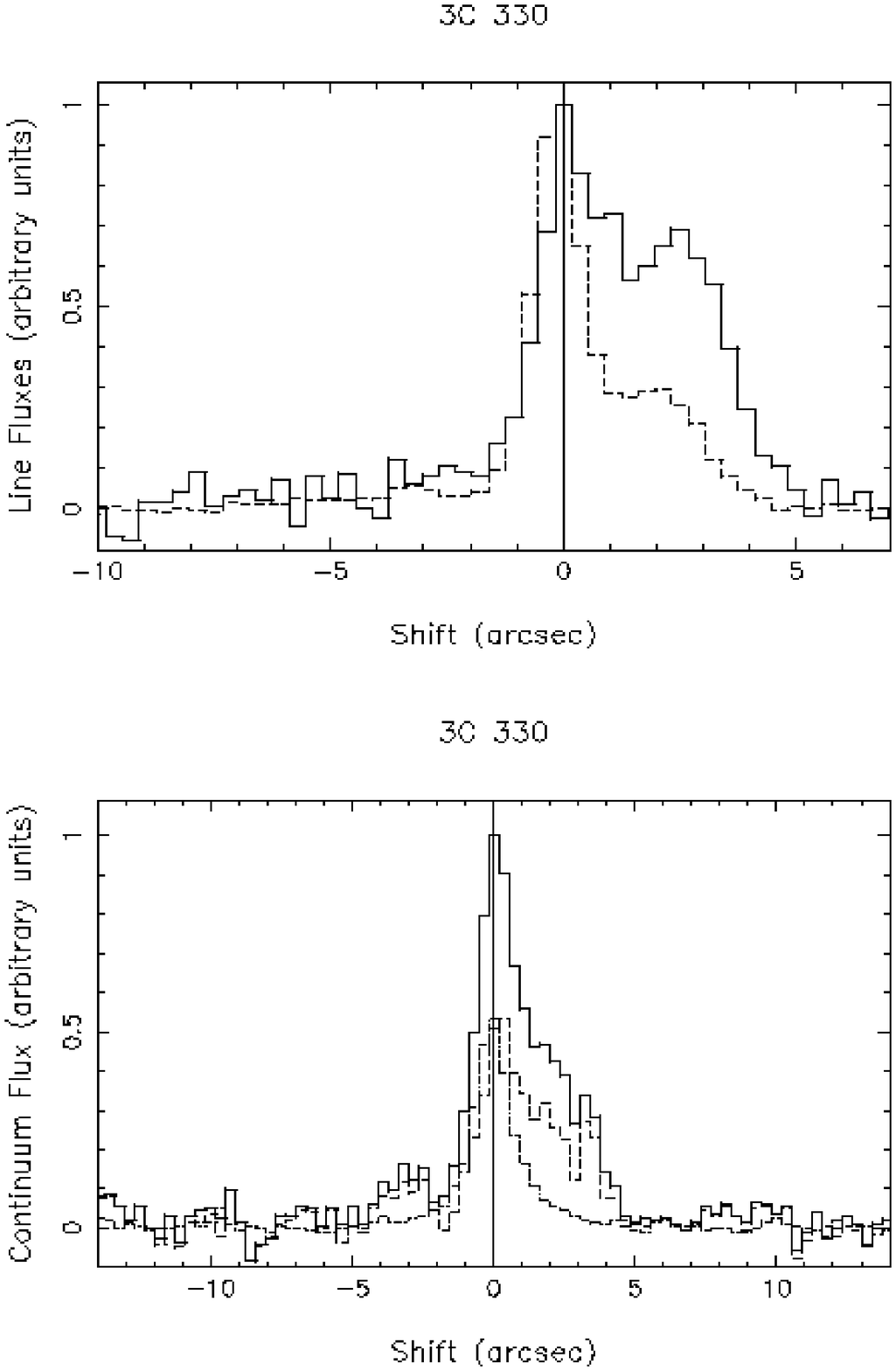,width=8.5cm}
\caption{Top: Variation of [OII] (solid line) and [OIII] (dashed line) fluxes along the radio axis of 3C 330. Bottom: Profile of the line-free continuum 
emission along the radio axis of 3C 330. The solid line shows the 3200 \AA \ -- 3700~\AA \ profile, the dashed line shows the blue-continuum emission after 
the subtraction of the nebular contribution (calculated from the H$\beta$ flux along the slit), and the dotted line shows the 5200 \AA \ -- 5700 \AA \ profile 
(scaled to the peak of the nebular-subtracted blue profile). (North-East is to the right)}
\label{330linecont}
\end{figure}

Figure~\ref{330linecont} (top) shows the variation of the [OII]$\lambda3727$ (solid line) and [OIII]$\lambda5007$ (dashed line) fluxes along the radio axis of 3C 330. Since the radio hot spots of 3C 330 are located outside the EELR, they are not shown in the figures. 

Line emission is seen extending approximately 14 arcsec (103 kpc) along the radio axis, with a central nuclear peak and a separated peak at $\sim$ 2 arcsec to the north-east of the continuum centroid. It can be noticed that the off-nucleus peak is significantly larger for [OII] than for [OIII], indicating that the ionization state is much lower in the northern extended region than in the nucleus.
\subsubsection{The emission-line kinematics}
\begin{figure}
\psfig{figure=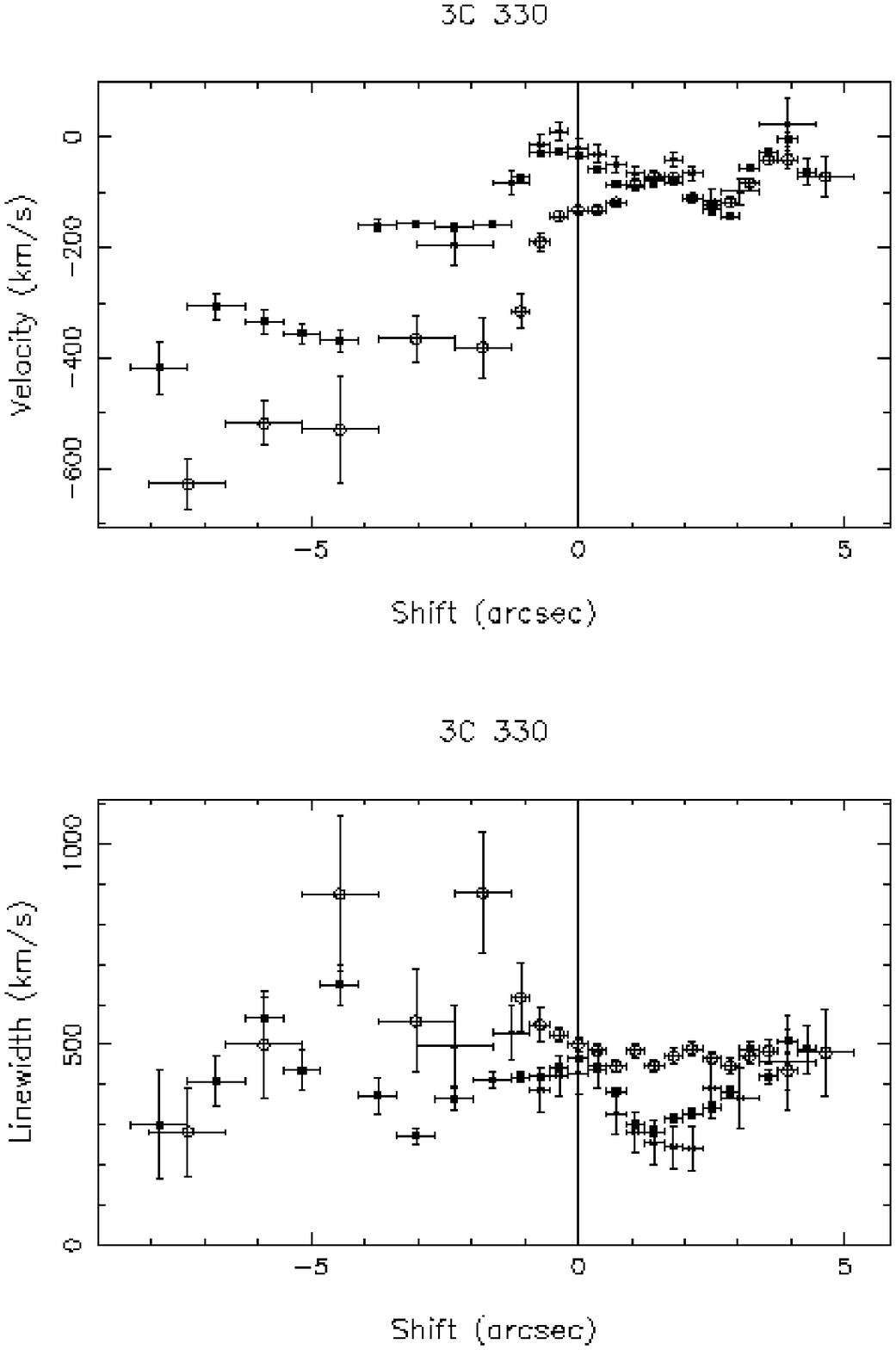,width=8.5cm}
\caption{Variation in the rest-frame radial velocity and in the linewidths for [OII] (open circles), H$\beta$ (asterisks) and [OIII] (filled squares) along 
the radio axis of 3C 330.}
\label{velwidth330}
\end{figure}

In contrast to the other ojects in our sample, the intermediate-resolution R300B grating was used for the blue observations of 3C 330, and the low-resolution R158R grating for the red observations, giving instrumental resolutions of 4.3 \AA \ and 7.4 \AA, respectively. Figure~\ref{velwidth330} (top) shows the variation in the velocity centroids of [OII]$\lambda3727$, H$\beta$ and [OIII]$\lambda5007$ along the radio axis of 3C 330. The [OII], H$\beta$ and [OIII] velocities vary in a similar manner along the northern extended region, from $\sim$ 1 to 5 arcsec. However, a clear contrast in the velocity centroids of [OII] with respect to [OIII] starts to appear in the nuclear region and reaches a maximum of $\sim$ 200 km s$^{-1}$ of difference between both velocities at the extreme of the southern EELR. Along this extended region, the [OII] velocity centroids follow a smooth curve, but the [OIII] velocity centroids vary in a chaotic way.
  
Figure~\ref{velwidth330} (bottom) shows how the linewidths of [OII]$\lambda3727$, H$\beta$ and [OIII]$\lambda5007$ vary along the radio axis of 3C 330. The low-ionization [OII] line is slightly broader than the high-ionization [OIII] line almost everywhere along the radio axis, except at both extremes of the slit where the errors are bigger and it is not so clear. However, the [OII] line was fitted with a single Gaussian, instead of fitting the doublet components separately; therefore, this could slightly broaden the measured [OII] linewidths (FWHM). In the nuclear and northern extended region, both [OII] and [OIII] linewidths follow smoooth curves, but they vary in a more chaotic way along the southern EELR. The H$\beta$ linewidth could only be measured in the central regions of the galaxy. In the nucleus and northern side of the nucleus, the linewidth of H$\beta$ varies in a similar manner as the [OIII] linewidth. However, on the southern side of the nucleus, it appears to be intermediate between those of the [OII] and [OIII] emission lines.

High-resolution long-slit spectra, taken at the same position angle (PA 230$^{o}$) by \scite{mccarthy96}, reveal that the [OIII] line emission extends $\sim$ 17 arcsec along the slit. The results they obtain in the kinematics are similar to ours, although they find a smooth velocity curve for the [OIII] emission line. The amplitude of both the velocity gradient and the linewidths for [OIII] are in agreement with what we obtain.

\begin{figure}
\psfig{figure=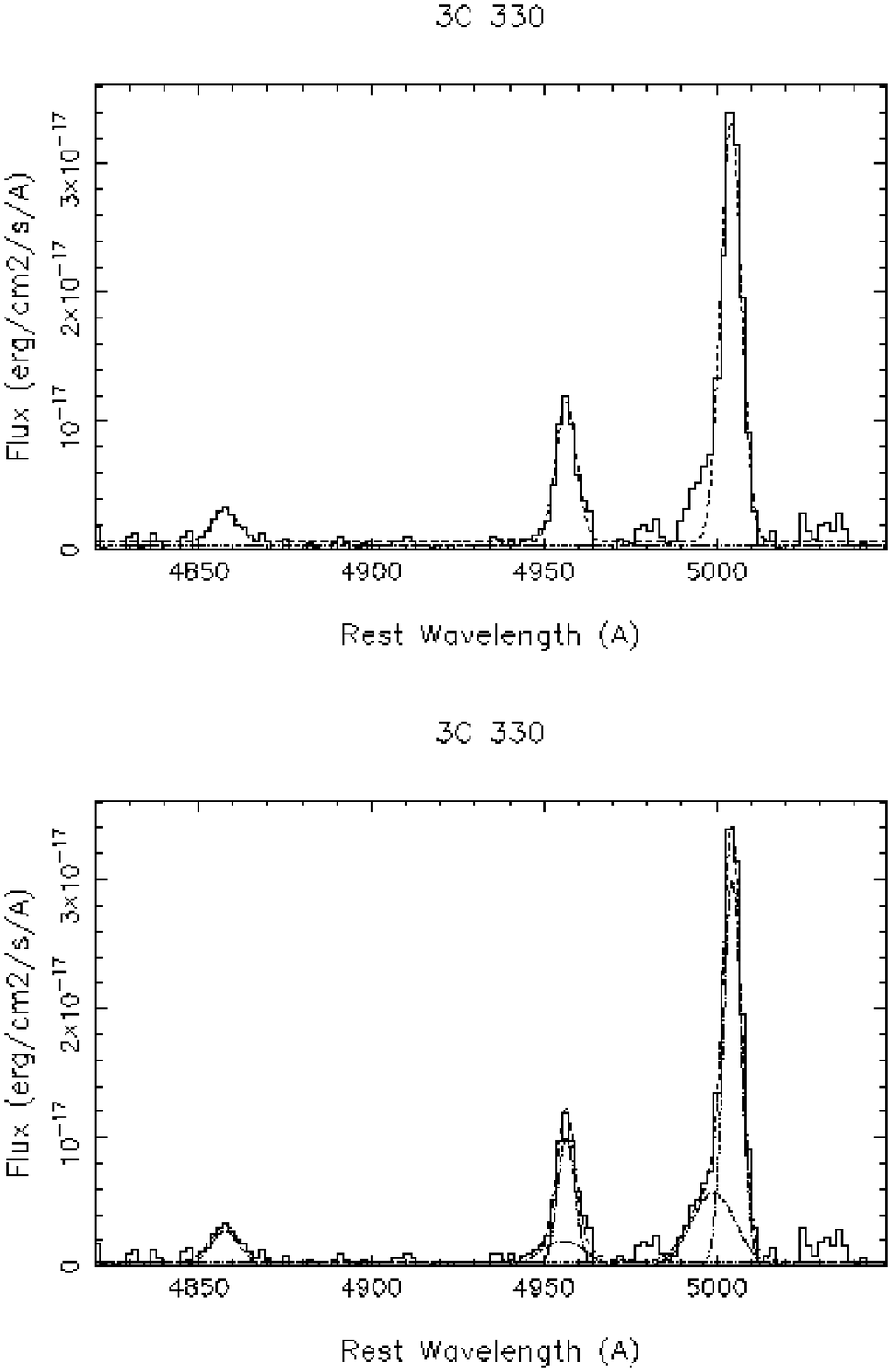,width=8.5cm}
\caption{H$\beta$, [OIII]$\lambda$4959 and [OIII]$\lambda$5007 emission-line profiles (solid line) from the red spectrum of 3C 330: 2.14 x 1.35 arcsec$^{2}$ 
aperture centred at 3.39 arcsec south-west of the nucleus. Top: The dashed line shows the simple Gaussian fit. Bottom: The two-component Gaussian fits (broad 
and narrow components)  (dot-dash-dot-dash lines) and the total fit (dashed line) are also plotted.}
\label{fits330}
\end{figure}

Figure~\ref{fits330} (top) shows the H$\beta$, [OIII]$\lambda4959$ and [OIII]$\lambda5007$ emission-line profiles from the red spectrum of 3C 330, corresponding to a 2.14 x 1.35 arcsec$^{2}$ aperture centred at 3.39 arcsec south-west of the nucleus. Initially, a single Gaussian fit (dashed line) was considered, but it is clear that the [OIII]$\lambda5007$ profile cannot be reproduced by this simple fit. The broad wing seen in the [OIII]$\lambda5007$ line suggests the existence of a broad component, in addition to the narrow component, in the kinematic structure of 3C 330. The two-component Gaussian fits (dot-dash-dot lines) and the total fit (dashed line) are presented in Figure~\ref{fits330} (bottom). It can be noticed that the H$\beta$ profile is dominated by the emission of the broad component.

\begin{figure}
\psfig{figure=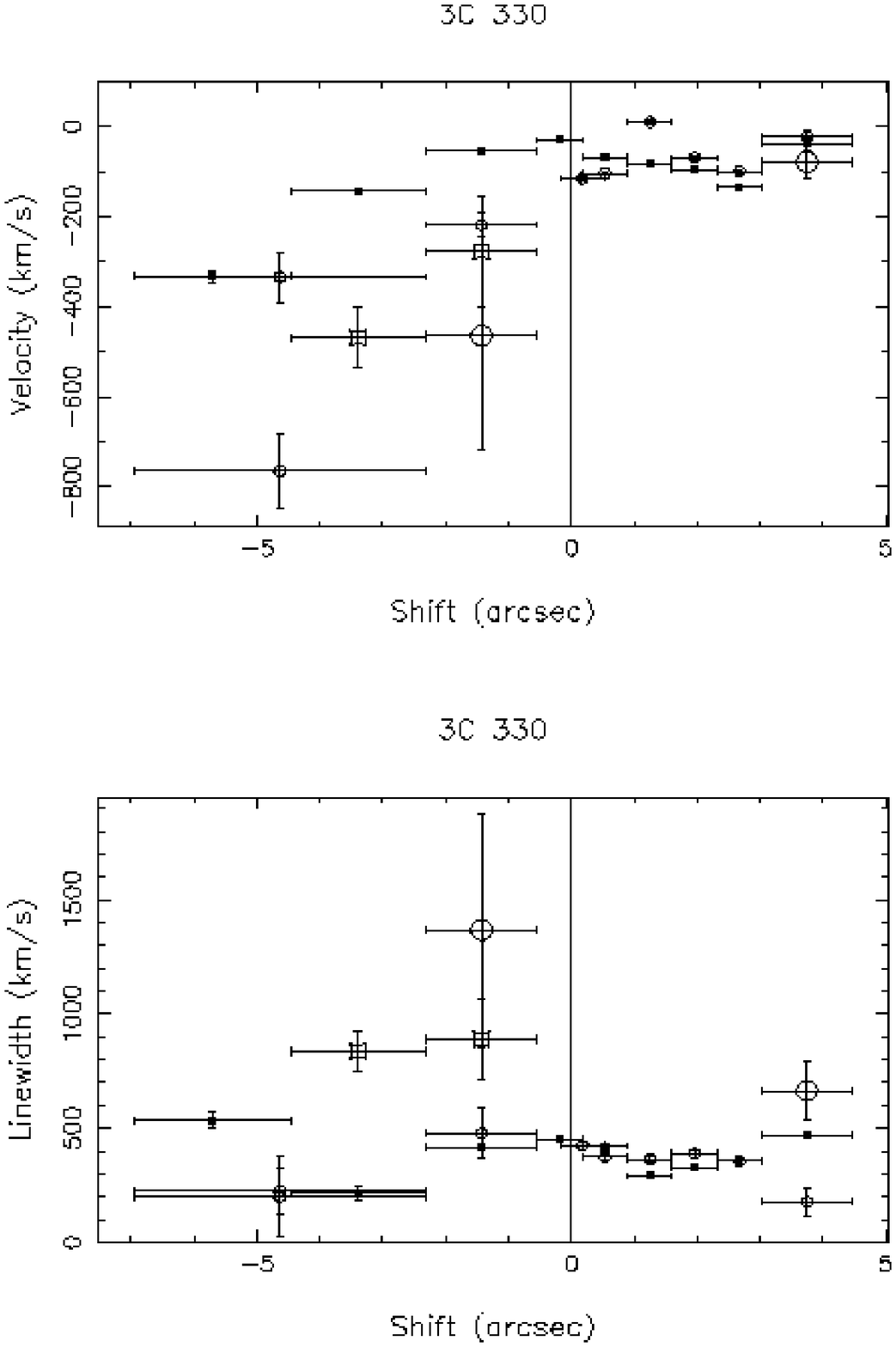,width=8.5cm}
\caption{Variation in the velocity centroids and in the linewidths (FWHM) along the radio axis of 3C 330 for the different kinematic components seen in [OII] 
(broad component (big open circles) and narrow component (small open circles)) and [OIII] (broad component (big open squares) and narrow component (small 
filled squares).}
\label{cvelwidth330}
\end{figure}

The variation in the velocity centroids along the radio axis of 3C 330 for the narrow and broad components seen in [OII]$\lambda3727$ and [OIII]$\lambda5007$ and the corresponding linewidths are shown in Figure~\ref{cvelwidth330}. Because of the high-resolution of the blue spectra, it was possible to fit each component of the [OII] doublet separately, in each kinematic component, assuming the low-density limit for the doublet ratio. A splitting in the [OII] emission line is detected at a radial distance of 2.5 to 7 arcsec (18 to 51 kpc) south-west of the nucleus. Note that the velocity centroids of the narrow components vary in a similar way, following a smooth curve, for [OII] and [OIII] in the southern region; in contrast, the broad components have a more distorted behaviour, giving the shift to the total velocities shown in Figure~\ref{velwidth330} (top).
\subsubsection{The emission-line spectra}
\begin{figure*}  
\begin{minipage}{160mm}
\psfig{figure=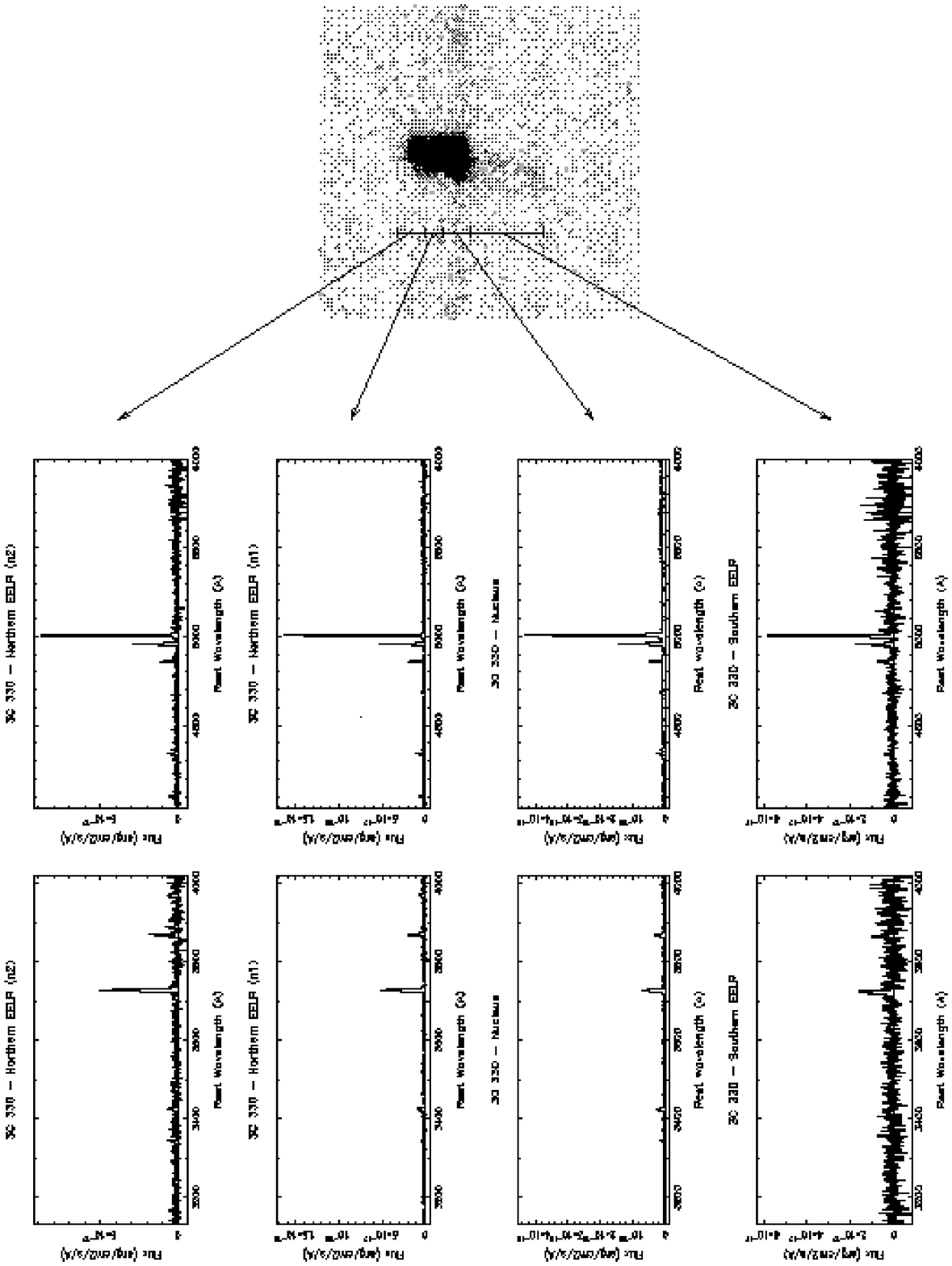,width=16cm}
\caption{Integrated blue and red spectra of the different regions in 3C 330, and 2-D spectrum showing [OII]$\lambda$3727 (3685 \AA \ -- 3765 \AA, 25.7 
arcsec along the slit).}
\label{330specim}
\end{minipage}
\end{figure*}

Two-dimensional spectrum showing the [OII]$\lambda$3727 emission line, and integrated blue and red spectra of four spatial regions (southern EELR, nucleus, n1- and n2-northern EELR) along the radio axis of 3C 330 are presented in Figure~\ref{330specim}. The integrated fluxes of the observed emission lines in these regions are given in Table~\ref{tabflux330}. The fluxes are normalised to the observed-frame H$\beta$ flux for each region. Since the Balmer-line intensities (H$\delta$ and H$\gamma$) relative to H$\beta$ are not significantly lower than expected in Case B recombination \cite{osterbrock89} in the extended emission-line regions, the off-nuclear line ratios were not corrected for intrinsic reddening. However, there is evidence for significant reddening in the nuclear aperture with both H$\delta$ and H$\gamma$ significantly less than expected for Case B recombination (E$_{B-V}=0.59\pm0.13$). The extinction implied by these results (A$_{B}=2.4\pm0.5$) is consistent with that inferred for the nuclear region of low-redshift radio galaxies (\pcite{tadhunter94}; \pcite{robinson2000}).

\begin{figure}  
\psfig{figure=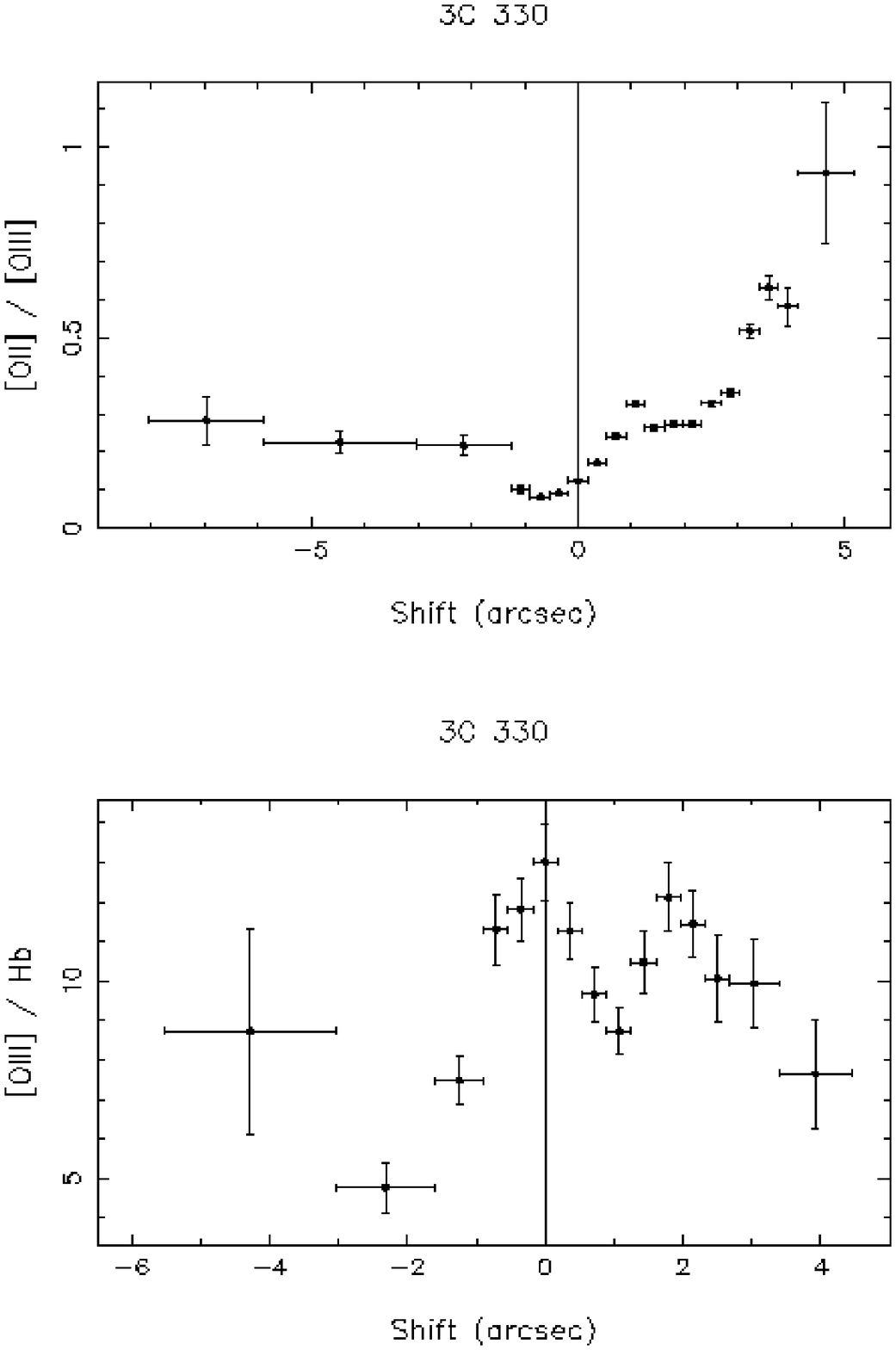,width=8.5cm}
\caption{Variation in the line ratios [OII]/[OIII] and [OIII]/H$\beta$ along the radio axis of 3C 330.}
\label{330ratios}
\end{figure}

Figure~\ref{330ratios}~shows~the~variations~in~the [OII]$\lambda3727$/[OIII]$\lambda5007$ and [OIII]$\lambda5007$/H$\beta$~line~ratios along the radio axis of 3C 330. It can be seen that the nuclear region is of high ionization with 10 $<$ [OIII]/H$\beta$ $<$ 13. The ionization state decreases to the northern side of the nucleus up to 1 arcsec, then rises up to $\sim$ 2 arcsec and decreases again towards the extreme of the northern EELR. On the southern EELR the [OII]/[OIII] line ratio shows little significant variation, but has a higher ionization state than the northern EELR. Unfortunately, only one point could be obtained along the southern region for the [OIII]/H$\beta$ line ratio.
\subsubsection{The physical conditions}
The electron temperature T$_{e}$ was measured for the central region and the EELR of 3C~330, by using the [OIII](4959+5007)/4363 line ratio. The [OIII]$\lambda$4363 linewidth was constrained to be the same as that of [OIII]$\lambda$5007. The values obtained from the red spectra are [OIII](4959+5007)/4363 = 33$\pm$7 for the southern EELR, 51$\pm$4 for the nucleus (after reddening correction), 66$\pm$8 for the n1-northern EELR and 44$\pm$7 for n2-northern EELR, which implies electron temperatures of T$_{e}$ = 23000$^{+5000}_{-3000}$ K, 17600$^{+500}_{-1000}$ K, 15300$^{+900}_{-800}$ K and 19000$^{+2000}_{-1500}$ K, respectively. A density of n$_{e}$ = 100 cm$^{-3}$ was assumed.
\subsubsection{The continuum emission}
Figure~\ref{330linecont} (bottom) shows spatial profile of the continuum emission along the radio axis of 3C 330. The solid line represents the profile of the UV continuum emission (3200 \AA \ -- 3700 \AA), the dashed line represents the UV continuum after the subtraction of the nebular continuum (calculated from the rest-frame H$\beta$ flux along the slit, in the same way as that of 3C 34, by using the {\small NEBCONT} routine in DIPSO (see Section~\ref{nebcont})), and the dotted line represents the green continuum emission (5200 \AA \ -- 5700 \AA), scaled to the peak of the nebular-subtracted UV profile. A UV excess can be observed approximately between 2 and 4 arcsec north-east of the nucleus. It is found that the contribution of the nebular emission is $\sim$ 50 per cent of the UV continuum emission in the nucleus of 3C 330, and $\sim$~35~per cent in the northern EELR, for the wavelength range 3200 \AA \ -- 3700 \AA. However, the nebular continuum cannot account for the observed UV excess at the north-east side of the nucleus, since this excess still remains after the nebular subtraction.
\subsection{3C 352}
\subsubsection{Previous observations}
3C 352 is a compact 10.2 arcsec (84 kpc) double radio source at redshift z = 0.8067 \cite{mccarthy95}. It was first studied optically by \scite{smith79} 
who found a low-ionization emission-line spectrum. \scite{hippelein92} present Fabry-P\'{e}rot imaging interferometry of 3C 352, and conclude that the morphology and velocity structure of the extended emission-line region around 3C 352 can be explained in terms of cooling gas ionized and accelerated by the bowshock associated with the radio jet, while some ionization close to the centre could also be due to radiative ionization. Based on long-slit spectroscopy observations, \scite{best2000a} study the UV emission-line ratios [CIII]$\lambda$2326/[CII]$\lambda$1909 and [NeIII]$\lambda$3869/[NeV]$\lambda$3426 of 3C 352, and compare them with theoretical predictions of shock and photoionization models, finding that the line ratios are in agreement with shock models including a precursor region. 
\subsubsection{The emission-line structure}
\begin{figure}
\psfig{figure=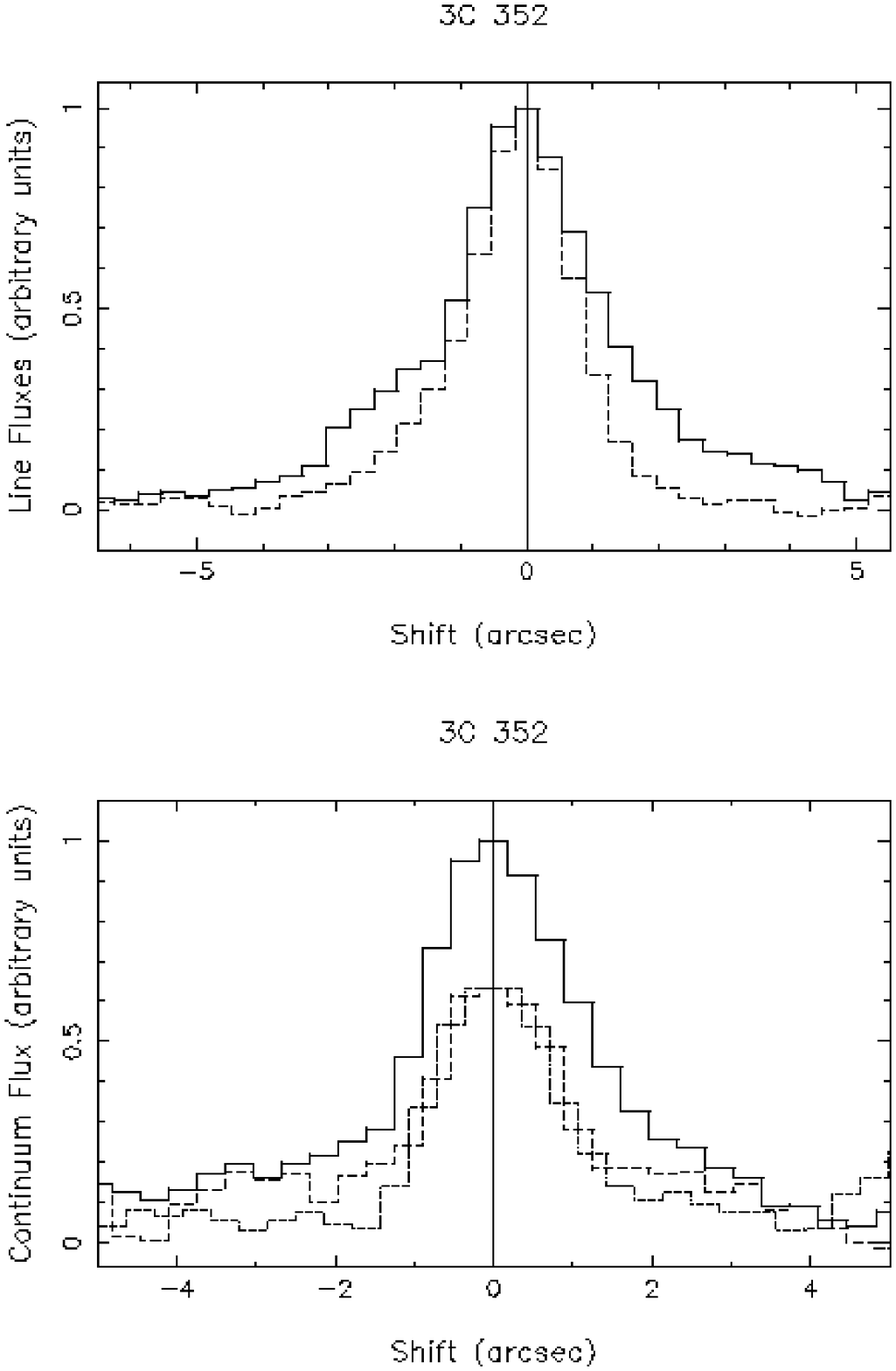,width=8.5cm}
\caption{Top: Variation of [OII] (solid line) and [OIII] (dashed line) fluxes along the radio axis of 3C 352. Bottom: Profile of the line-free continuum 
emission along the radio axis of 3C 352. The solid line shows the 2600 \AA \ -- 3700~\AA \ profile, the dashed line shows the blue-continuum emission after 
the subtraction of the nebular contribution (calculated from the H$\beta$ flux along the slit), and the dotted line shows the 4400 \AA \ -- 4800 \AA \ profile 
(scaled to the peak of the nebular-subtracted blue profile). (North-West is to the right)}
\label{352linecont}
\end{figure}

Figure~\ref{352linecont} (top) presents the [OII]$\lambda3727$ (solid line) and [OIII]$\lambda5007$ (dashed line) spatial profiles along the radio axis of 3C 352. They were derived by extracting spatial slices, following the same procedure of the other galaxies (see Section~\ref{contsubtraction}). It can be seen that the [OII] line emission extends $\sim$ 10 arcsec (82 kpc) along the radio axis, while the [OIII] line emission only extends $\sim$ 6 -- 7 arcsec (53 kpc) along the same direction. The radio hot spots in this radio galaxy lie just outside the EELR, and they are not shown in the figures. 
\subsubsection{The emission-line kinematics}
\begin{figure}
\psfig{figure=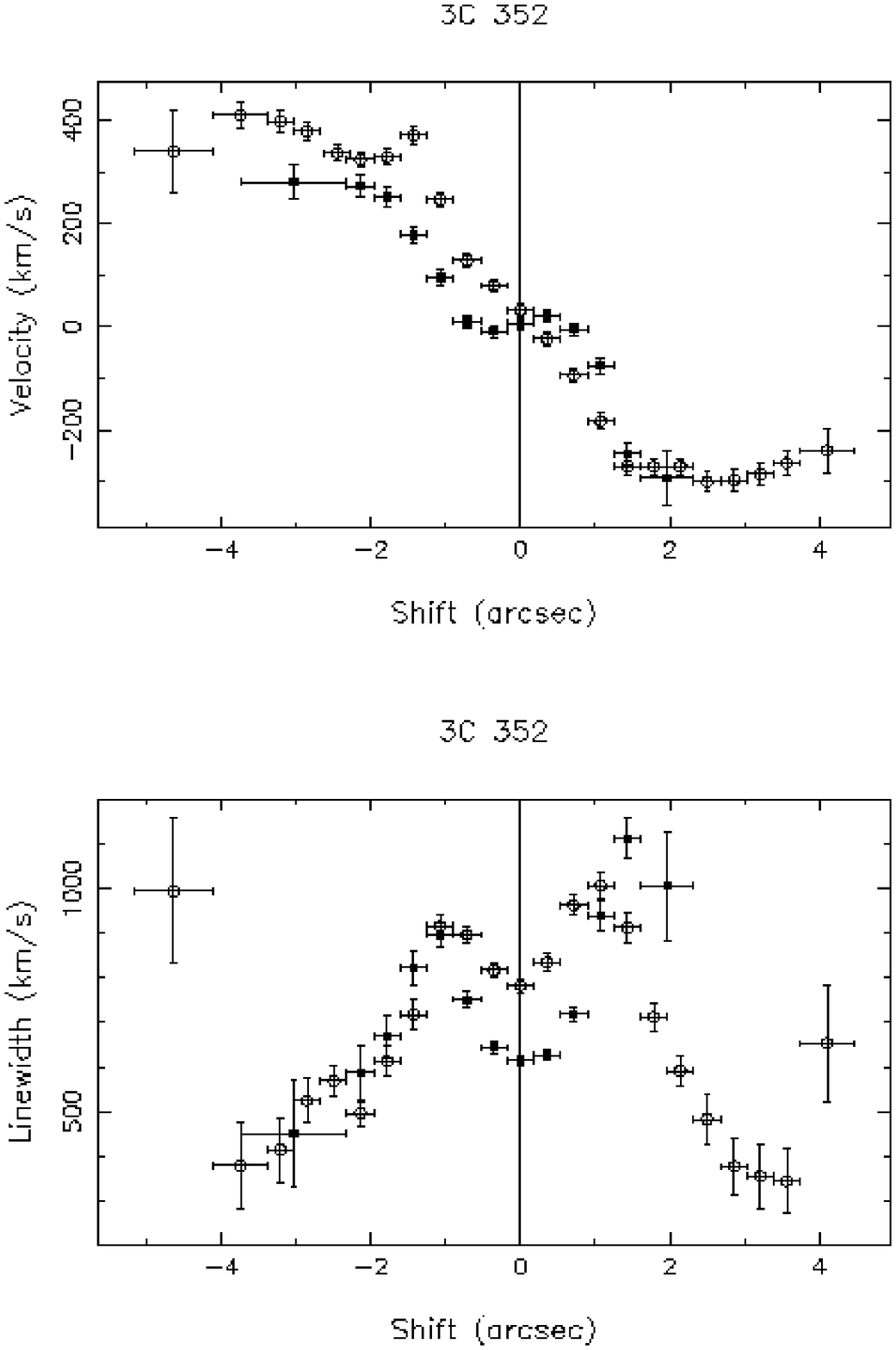,width=8.5cm}
\caption{Variation in the rest-frame radial velocity and in the linewidths along the radio axis of 3C 352 for [OII] (open circles) and [OIII] (filled 
squares).}
\label{velwidth352}
\end{figure}

Figure~\ref{velwidth352} (top) shows how the velocity centroids of [OII]$\lambda3727$ and [OIII]$\lambda5007$ vary along the radio axis of 3C 352. They follow a smooth rotation curve and both vary in a similar manner. 

The variation in the linewidths of [OII]$\lambda3727$ and [OIII]$\lambda5007$ is presented in Figure~\ref{velwidth352} (bottom). Both [OII] and [OIII] linewidths vary in a similar way. However, in the central region it appears that the low-ionization [OII] line is broader than the high-ionization [OIII] line. From this figure, it can also be noticed that the [OIII] line emission does not extend as far as the [OII] line emission along the radio axis.

\begin{figure}
\psfig{figure=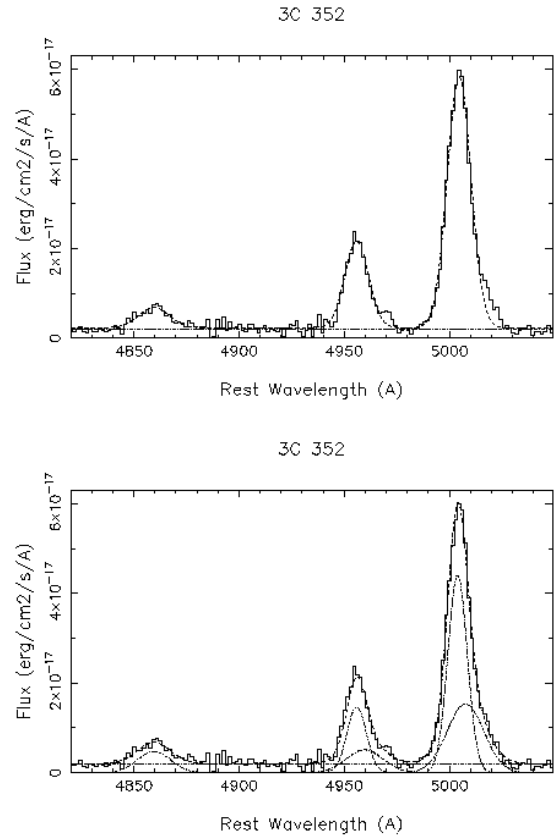,width=8.5cm}
\caption{H$\beta$, [OIII]$\lambda$4959 and [OIII]$\lambda$5007 emission-line profiles (solid line) from the red spectrum of 3C 352: 1.07 x 1.56 arcsec$^{2}$ 
aperture centred at 0.71 arcsec south-east of the nucleus. Top: The dashed line shows the simple Gaussian fit. Bottom: The two-component Gaussian fits (broad 
and narrow component) (dot-dash-dot-dash lines) and the total fit (dashed line) are also plotted.}
\label{fits352}
\end{figure}

Figure~\ref{fits352} shows the H$\beta$, [OIII]$\lambda4959$ and [OIII]$\lambda5007$ emission-line profiles from the red spectrum of 3C 352, corresponding to a 1.07 x 1.56 arcsec$^{2}$ aperture centred at 0.71 arcsec south-east of the nucleus. A single Gaussian fit, considering line emission from only one kinematical component, is shown at the top. However, the broad wings in [OIII] suggest the existence of a broad component in the kinematic structure of 3C 352. The two-component fits (dot-dash-dot lines) and the resulting fit (dashed line) are presented at the bottom of Figure~\ref{fits352}. It can be seen that the broad component dominates the emission of H$\beta$.

\begin{figure}
\psfig{figure=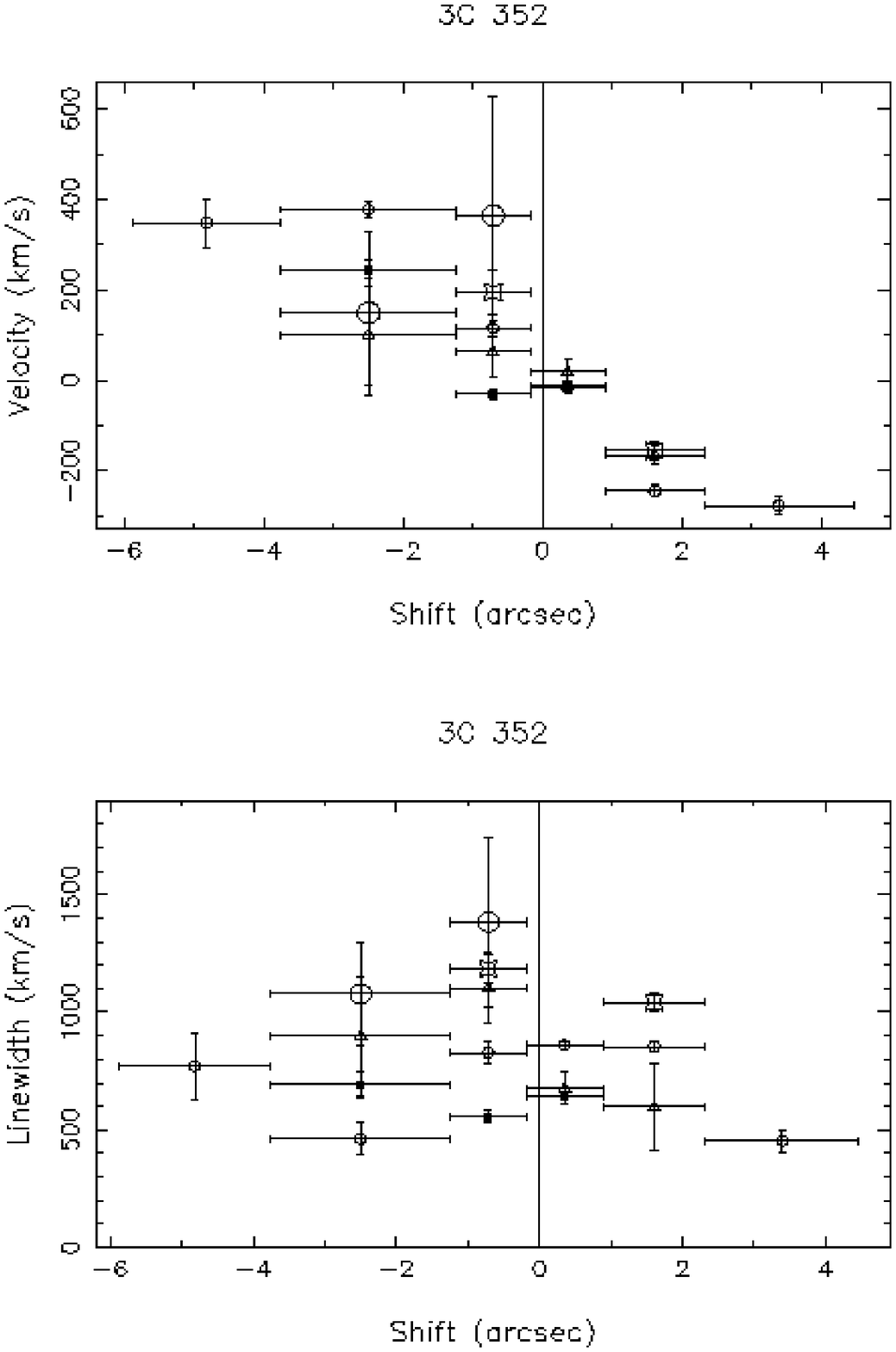,width=8.5cm}
\caption{Variation in the velocity centroids and in the linewidths (FWHM) along the radio axis of 3C 352 for the different kinematic components seen in [OII] 
(broad component (big open circles) and narrower component (small open circles)), H$\beta$ (open triangles) and [OIII] (broad component (big open squares) and 
narrower component (small filled squares)).}
\label{cvelwidth352}
\end{figure}

Figure~\ref{cvelwidth352} (top) shows the variation of the velocity centroids along the radio axis of 3C 352 for the broad and narrow components seen in [OII]$\lambda3727$ and [OIII]$\lambda5007$. The velocity centroids of H$\beta$ are also plotted. The variation in the linewidth along the radio axis for the two kinematical components seen in [OII]$\lambda3727$ and [OIII]$\lambda5007$ is shown in Figure~\ref{cvelwidth352} (bottom) together with the linewidth of H$\beta$.
\subsubsection{The emission-line spectra}
\begin{figure*}
\begin{minipage}{160mm}
\psfig{figure=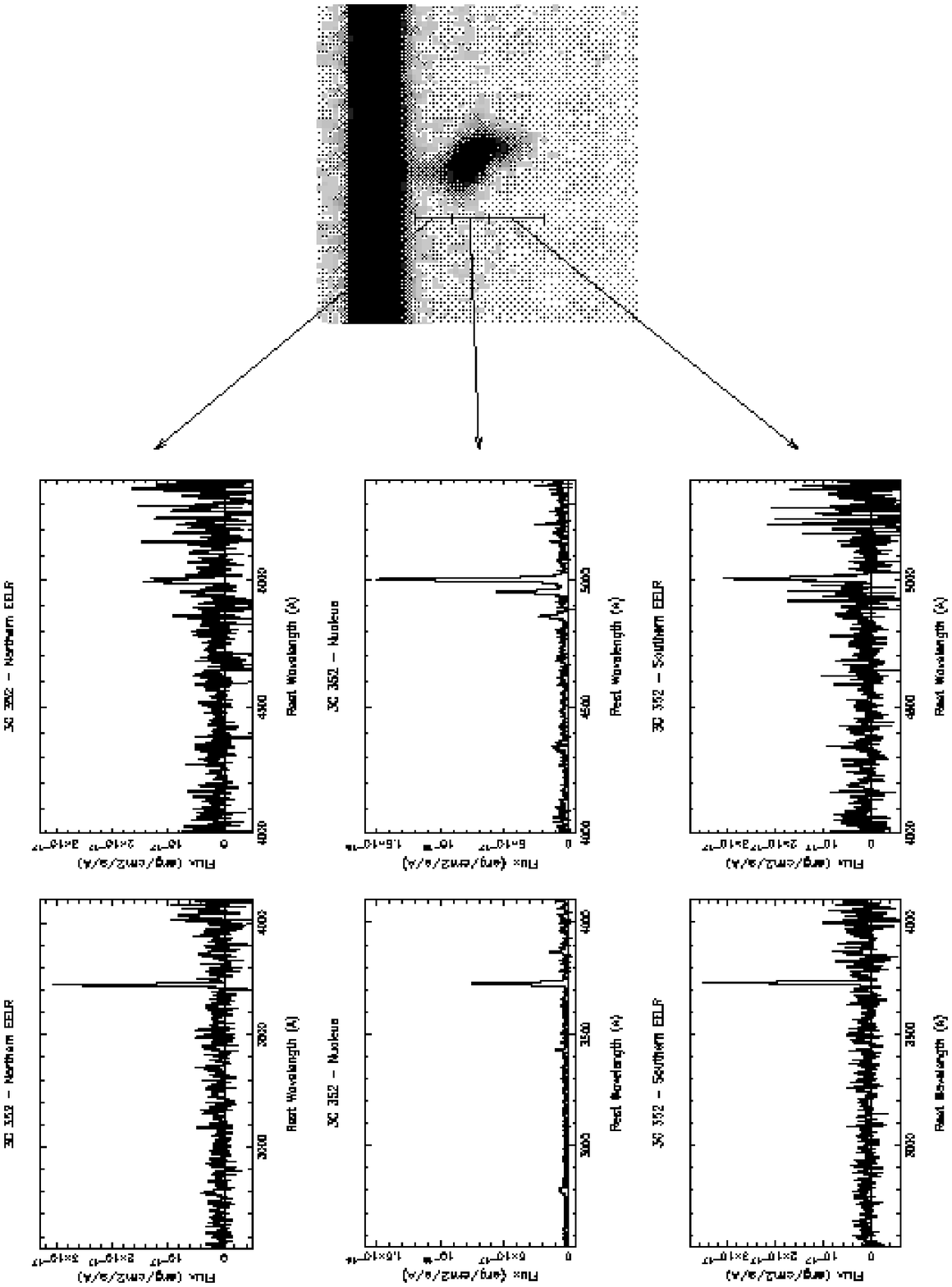,width=16cm}
\caption{Integrated blue and red spectra of the different regions in 3C 352, and 2-D spectrum showing [OII]$\lambda$3727 (3655 \AA \ -- 3795 \AA, 28.2 
arcsec along the slit).}
\label{352specim}
\end{minipage}
\end{figure*}

Figure~\ref{352specim} shows integrated blue and red spectra of the central nuclear region, and the southern and northern EELR of 3C 352, a 2-D spectrum showing the [OII]$\lambda$3727 emission line is also plotted. Fluxes of the emission lines in the three regions are listed in Table~\ref{tabflux352}, and are normalised to the observed-frame H$\beta$ flux in each region. Given that the H$\delta$ and H$\gamma$ fluxes relative to H$\beta$ are consistent with Case B recombination \cite{osterbrock89}, no corrections for intrinsic reddening have been made to the line fluxes.
 
\begin{figure}
\psfig{figure=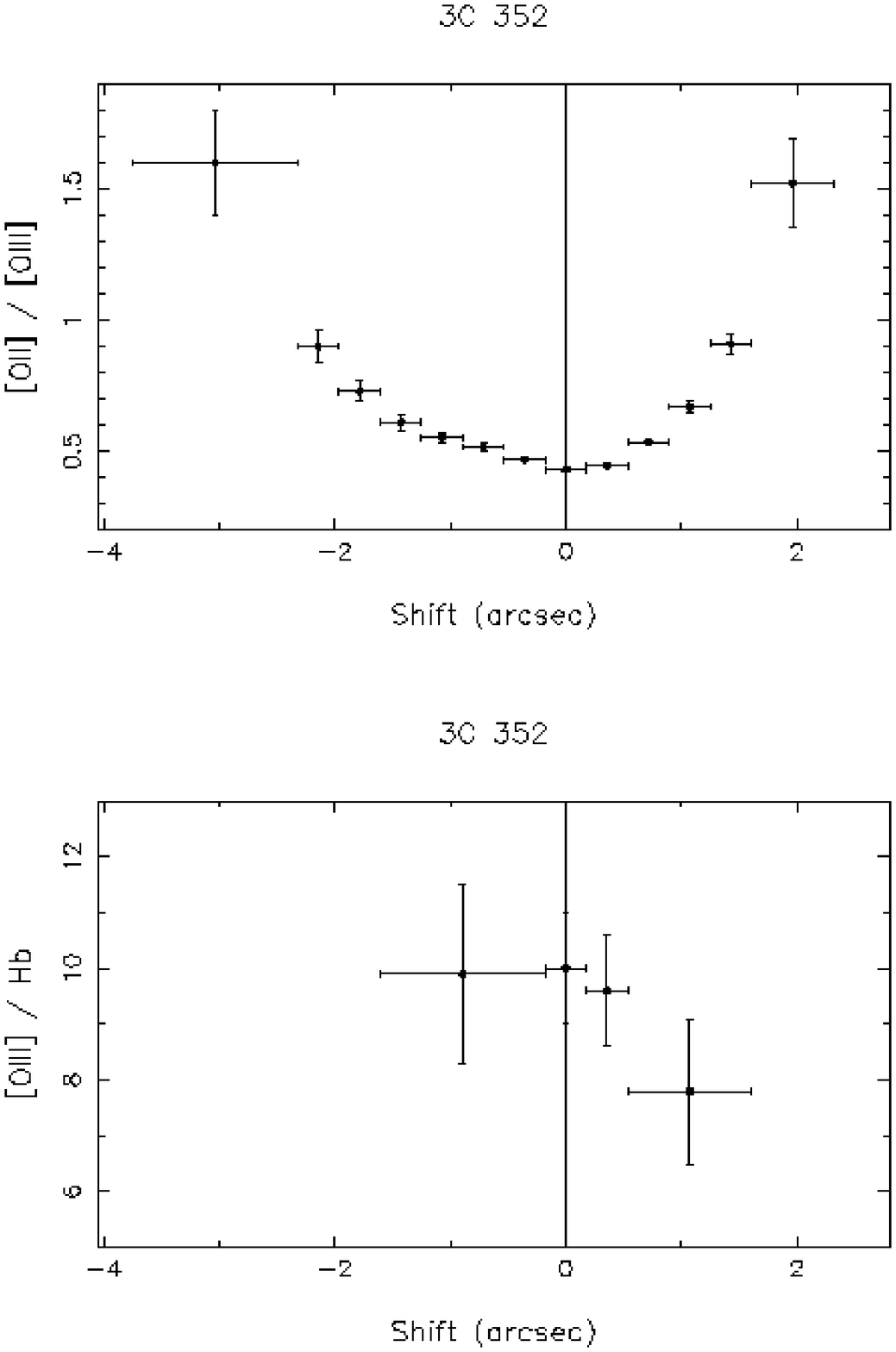,width=8.5cm}
\caption{Variation in the line ratios [OII]/[OIII] and [OIII]/H$\beta$ along the radio axis of 3C 352.}
\label{352ratios}
\end{figure}

The variation in the [OII]$\lambda3727$ / [OIII]$\lambda5007$ along the radio axis is shown in Figure~\ref{352ratios} (top). It can be observed that the ionization state peaks where the continuum centroid is situated, and then falls on either side of the nucleus, more steeply towards the northern side, as the radius increases.

Figure~\ref{352ratios} (bottom) shows the [OIII]$\lambda5007$ / H$\beta$ line ratio along the radio axis of 3C 352. Unfortunately, the H$\beta$ line could only be measured in the nuclear region. However, from this figure it can be noticed that the nucleus has a high ionization state ([OIII]/H$\beta$ $\sim$ 10).
\subsubsection{The physical conditions}
The~electron~temperature~T$_{e}$~was~measured~for~the central \ nuclear region of 3C 352 by using the [OIII](4959+5007)/4363 line ratio; but this measurement was not possible for the two EELR. The value obtained for the nucleus from the red data is [OIII](4959+5007)/4363 = 46$\pm$7, which implies a temperature T$_{e}$ = 18000$^{+2000}_{-1000}$ K for the nuclear region. A density of n$_{e}$ = 100 cm$^{-3}$ was assumed.
\subsubsection{The continuum emission}
The spatial profile of the continuum emission along the radio axis of 3C 352 is shown in Figure~\ref{352linecont} (bottom). The solid line shows the UV (2600 \AA \-- 3700 \AA) continuum profile, the dashed line shows the UV continuum after the subtraction of the nebular continuum contribution (derived from the rest-frame H$\beta$ flux along the slit, by using the {\small NEBCONT} routine in DIPSO (see Section~\ref{nebcont})), and the dotted line shows the 4400 \AA \ -- 4800 \AA \ continuum profile, scaled to the peak of the nebular-subtracted UV profile. The emission from the star situated just about 6.5 arcsec north-west of 3C 352 was subtracted before extracting the continuum profile of the radio galaxy. It is found that the nebular emission contribution with respect to the continuum emission is $\sim$ 37 per cent in the nucleus and in the southern EELR, and $\sim$ 28 per cent in the northern EELR, for the wavelength range 2600 \AA \ -- 3700 \AA. There is no evidence for excess UV continuum following the subtraction of the nebular continuum: the nebular-subtracted UV and longer wavelength continuum spatial profile are not significantly different (although the presence of the nearby bright star makes the comparison difficult). This behaviour is similar to that observed in the low-redshift jet-cloud interaction 3C 171 \cite{clark96}.
\subsection{3C 435A}
\subsubsection{Previous observations}
The radio source 3C 435A, at a redshift z =  0.471 and with an angular size of 14 arcsec (97 kpc), was discovered at a small projected distance from the powerful radio source 3C 435B \cite{mccarthy89}, but the two objects are at different redshifts. Narrow-band [OII] emission-line imaging shows presence of an extended emission-line nebula that extends well beyond the compact radio source on both sides of the nucleus (van Breugel \& McCarthy 1989). Bidimensional spectroscopy has been performed with the integral field spectrograph TIGER \cite{roccavol94}, showing at least five components around the central one. A scenario of gas being overpressured by radio plasma expansion, or by the interaction of relativistic electrons with the ambient gas, was favoured. Kinematic analysis by \scite{mccarthy96}, based on long-slit spectroscopy along PA 229$^{o}$, shows a complex velocity profile, with an overall amplitude of a few hundred km s$^{-1}$, and apparent rotation in both the central galaxy and in the component lying $\sim$ 4 arcsec north-east.
\subsubsection{The emission-line structure}
\begin{figure}
\psfig{figure=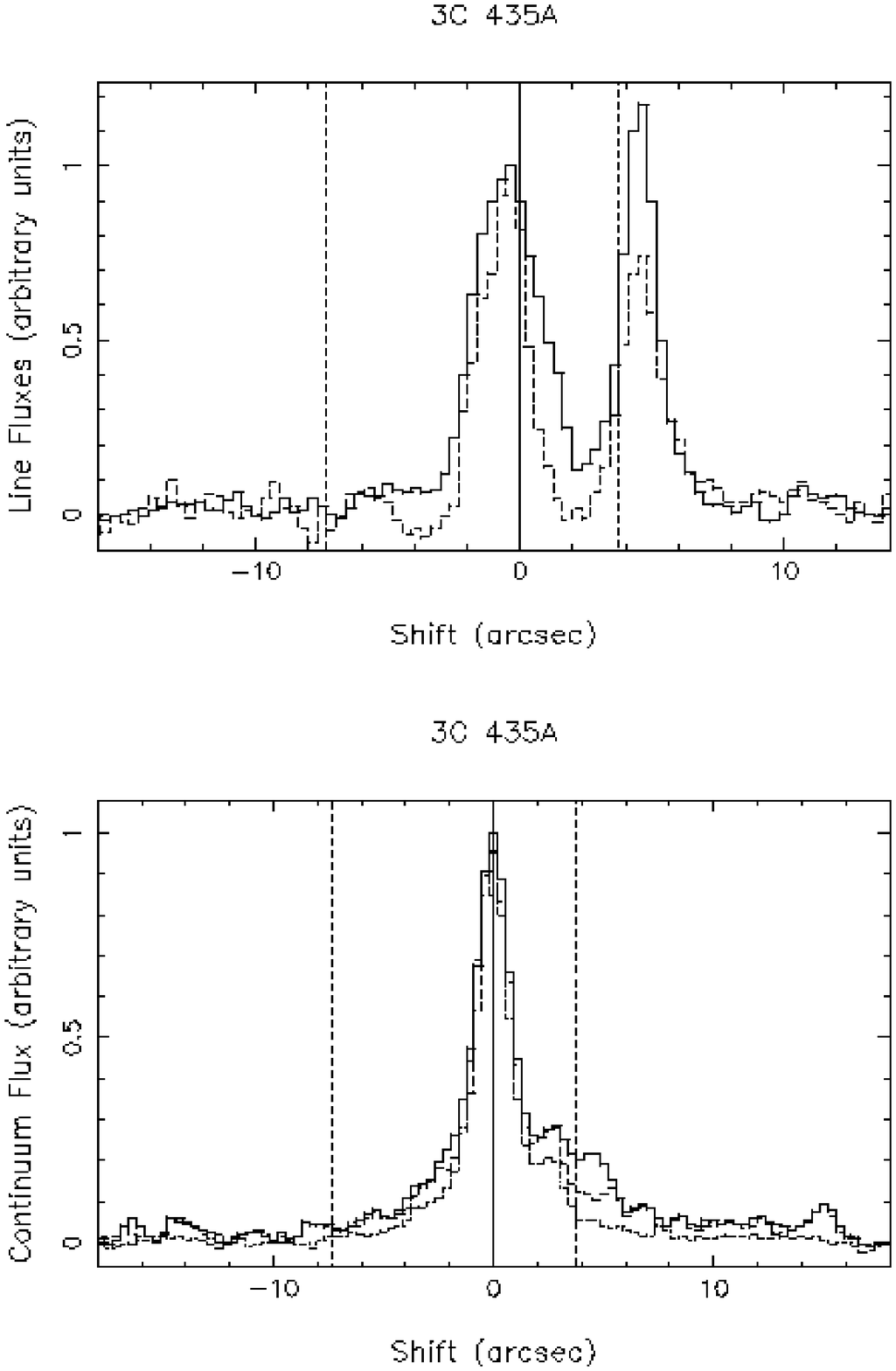,width=8.5cm}
\caption{Top: Variation of [OII] (solid line) and [OIII] (dashed line) fluxes along the radio axis of 3C 435A. Bottom: Profile of the line-free continuum emission along the radio axis of 3C 435A. The solid line shows the 2600 \AA \ -- 3700~\AA \ profile, the dashed line shows the blue-continuum emission after the subtraction of the nebular contribution (calculated from the H$\beta$ flux along the slit), and the dotted line shows the 5200 \AA \ -- 6000 \AA \ profile (scaled to the peak of the nebular-subtracted blue profile). (North-East is to the right)}
\label{435alinecont}
\end{figure}

The spatial profiles of the [OII]$\lambda3727$ (solid line) and [OIII]$\lambda5007$ (dashed line) emission lines are presented in Figure~\ref{435alinecont} (top). These profiles were obtained in the same way as those of the other four radio galaxies, by extracting spatial slices through the long-slit spectra (see Section~\ref{contsubtraction}). 

The projected positions of the radio hot spots are indicated on the plots by dashed vertical lines. Line emission extends approximately 24 arcsec (166 kpc) along the radio axis. In addition to the central galaxy, a companion can be seen located at $\sim$ 4 -- 5 arcsec (30 kpc) north-east of the continuum centroid along the radio axis (see also \pcite{mccarthy89}). The companion is situated in the interaction zone of the radio jet with the intergalactic medium \cite{roccavol94}. The fact that the [OII] line is relatively stronger in this component indicates that it has lower ionization state than the central component.
\subsubsection{The emission-line kinematics}
Since the slit positions and the grating angles were coincident for the red spectra taken on two different nights, the individual frames were coadded to produce a single long-slit spectra, and therefore increasing the S/N ratio. 

\begin{figure}
\psfig{figure=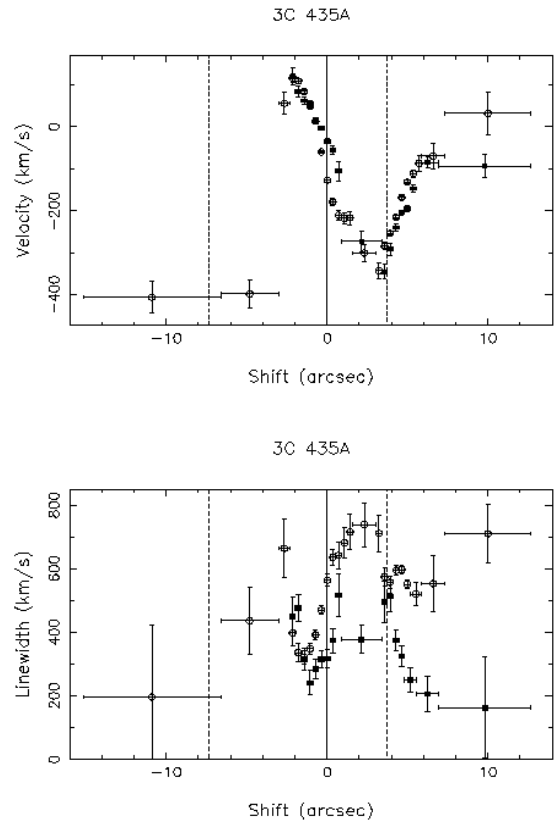,width=8.5cm}
\caption{Variation in the rest-frame radial velocity and in the linewidths (FWHM) of [OII] (open circles) and [OIII] (filled squares) along the radio axis of 3C 435A.}
\label{velwidth435a}
\end{figure}

Figure~\ref{velwidth435a} (top) shows how the velocity centroids of [OII]$\lambda3727$ and [OIII]$\lambda5007$ vary along the radio axis of 3C 435A. Both [OII] and [OIII] velocities are seen to vary in a similar manner, with rotation curve-like velocity variations in both central and northern components. 

The variation in the linewidths of [OII]$\lambda3727$ and [OIII]$\lambda5007$ along the radio axis of 3C 435A is shown in Figure~\ref{velwidth435a} (bottom). The low-ionization [OII] line tends to be broader than the high-ionization [OIII] line, especially in the region between the nucleus and the northern radio hot spot. Also, in the northern EELR the [OIII] emission line seems to be much narrower than the [OII] emission line, \emph {beyond} the radio hot spot. In this case, the [OIII] linewidth might have been affected by the subtraction of the night-sky line that falls on it, since the S/N ratio in the EELR is much lower than in the central regions. However, the [OIII] linewidth results agree with those of \scite{mccarthy96}.

\begin{figure}
\psfig{figure=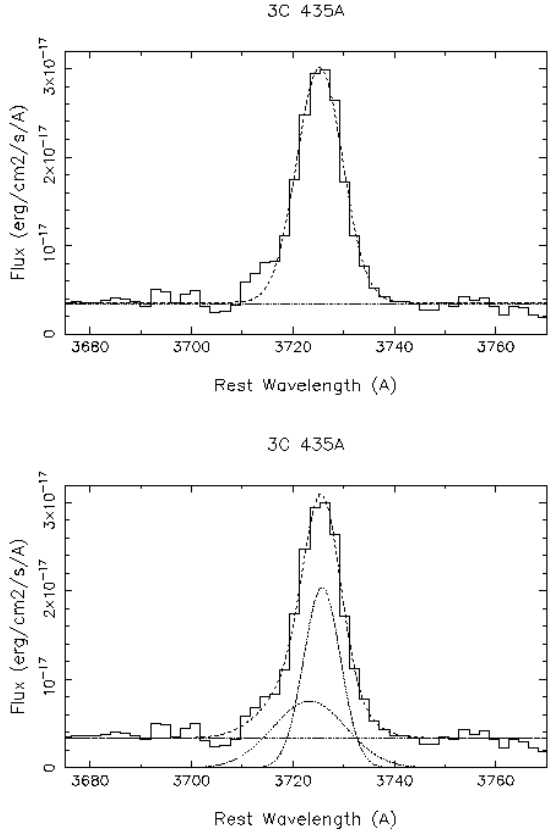,width=8.5cm}
\caption{[OII]$\lambda$3727 emission-line profile (solid line) from the blue spectrum of 3C 435A: 2.50 x 1.56 arcsec$^{2}$ aperture centred at 1.43 arcsec north-east of the nucleus. Top: The dashed line shows the simple Gaussian fit. Bottom: The two-component Gaussian fits (broad and narrow components) (dot-dash-dot-dash lines) and the total fit (dashed line) are also plotted.}
\label{fits435a}
\end{figure}

Figure~\ref{fits435a} shows the [OII]$\lambda3727$ emission-line profile of 3C 435A, corresponding to a 2.50 x 1.56 arcsec$^{2}$ aperture centred at 1.43 arcsec north-east of the continuum centroid. Initially, a one-component Gaussian fit was considered (top), but this simple fit cannot reproduce the broad wing of [OII], which can be seen as emission from a broad kinematic component. The plot at the bottom of Figure~\ref{fits435a} presents the two-component (narrow and broad) Gaussian fit (dot-dash-dot lines) and the total fit (dashed line), which reproduces the [OII] profile much better than the one-component fit.

\begin{figure}
\psfig{figure=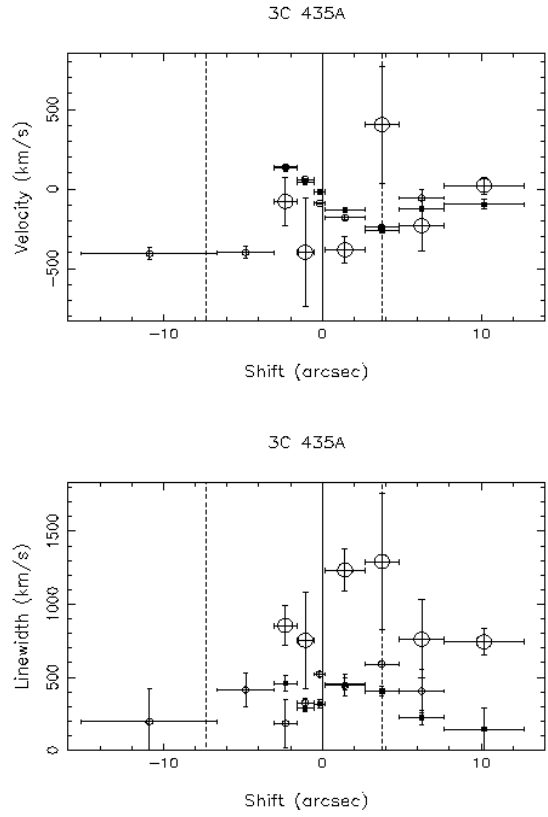,width=8.5cm}
\caption{Variation in the velocity centroids and in the linewidths (FWHM) along the radio axis of 3C 435A for the different kinematic components seen in 
[OII], broad component (big open circles) and narrow component (small open circles). For comparison, the variation in the velocity centroids and in the 
linewidth of [OIII] (filled squares) are also plotted.}
\label{cvelwidth435a}
\end{figure}

Figure~\ref{cvelwidth435a} (top) shows how the velocity centroids vary along the radio axis of 3C 435A for the broad and narrow components seen in [OII]$\lambda3727$. The [OIII]$\lambda5007$ velocity centroids are also plotted. No broad component could be fitted to the [OIII]$\lambda5007$ emission line at any spatial location. It can be seen that the narrow component of [OII] varies in a similar way as the [OIII] velocity centroids, consistent with rotation, while the broad component has a more distorted and chaotic behaviour. The variation in the linewidths along the radio axis for the two kinematic components seen in [OII] is presented in Figure~\ref{cvelwidth435a} (bottom). The linewidth of the broad component varies between $\sim$ 800 -- 1500 km s$^{-1}$ (FWHM), reaching the maximum values in the region between the continuum centroid and the northern radio hot spot.
\subsubsection{The emission-line spectra}
\begin{figure*}
\begin{minipage}{160mm}
\psfig{figure=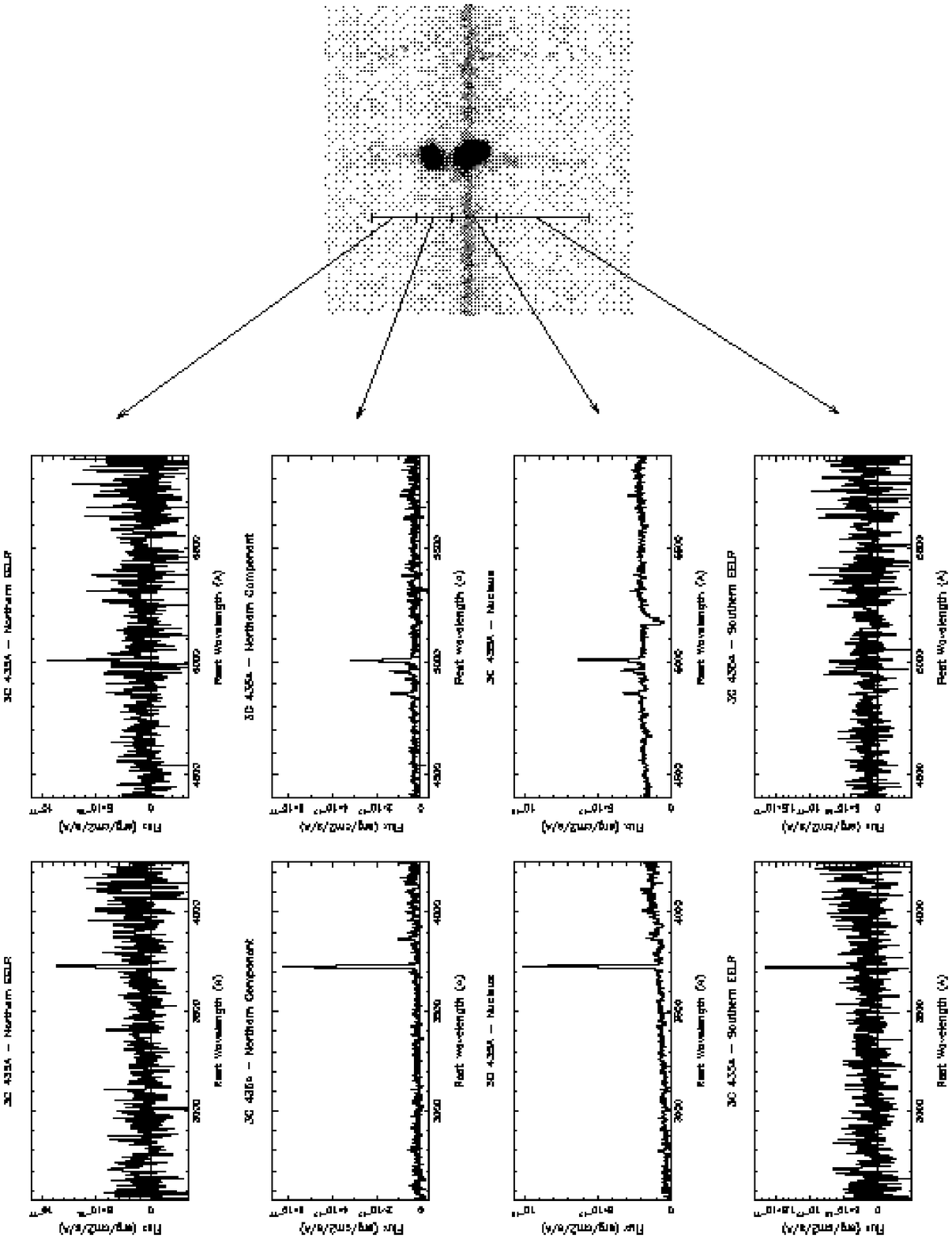,width=16cm}
\caption{Integrated blue and red spectra of the different regions in 3C 435A, and 2-D spectrum showing [OII]$\lambda$3727 (3625 \AA \ -- 3825 \AA, 38.5 
arcsec along the slit).}
\label{435aspecim}
\end{minipage}
\end{figure*}

Integrated blue and red spectra of the southern EELR, the central nuclear region, the northern component and the northern EELR of 3C 435A, and a 2-D spectrum showing the [OII]$\lambda$3727 emission line are presented in Figure~\ref{435aspecim}. The fluxes of the observed emission lines in the four regions are listed in Table~\ref{tabflux435a}, and are normalised to the observed-frame H$\beta$ flux for each region. The line fluxes have not been corrected for intrinsic reddening, given that the H$\alpha$/H$\beta$ ratio is not significantly larger than the theoretical value corresponding to Case B recombination \cite{osterbrock89}, suggesting that the intrinsic reddening is not large.

Figure~\ref{435aratios} shows the [OII]$\lambda3727$/[OIII]$\lambda5007$ and the [OIII]$\lambda5007$/H$\beta$ line ratios along the radio axis of 3C 435A. From these plots it can be noticed that the region between the nucleus and the northern radio hot spot has the lowest ionization state, 1$<$ [OIII]/H$\beta$ $<$ 2, although, the ionization state is seen to be low everywhere. It can also be seen that the nucleus has higher ionization state than the northern companion located $\sim$ 4 -- 5 arcsec to the north.
\subsubsection{The physical conditions}
For 3C 435A, the [OIII]$\lambda$4363 was not measured because it was outside the wavelength range covered. Therefore, the temperature diagnostic could not be obtained for this galaxy.
\subsubsection{The continuum emission}
The spatial profile of the continuum emission along the radio axis of 3C 435A is shown in Figure~\ref{435alinecont} (bottom). The solid line represents the UV (2600 \AA \ -- 3700 \AA) continuum emission, the dashed line represents the UV continuum after the subtraction of the nebular continuum contribution (calculated from the rest-frame H$\beta$ flux along the slit, by using the {\small NEBCONT} routine in DIPSO (see Section~\ref{nebcont})), and the dotted line represents the green (5200 \AA \ -- 6000 \AA) continuum profile, scaled to the peak of the nebular-subtracted UV profile. It is found that the nebular emission with respect to the continuum emission is $\sim$ 10 per cent in the nucleus, and $\sim$ 21 per cent at the location of the northern component, for the wavelength range 2600 \AA \ -- 3700 \AA. On the basis of the comparison between the nebular-subtracted UV and the red continuum slices, there is evidence for a UV excess on both sides of the nucleus, especially at the location of the northern hot spot. \emph{Note that the extended UV continuum emission (nebular-subtracted) extends over the entire emission line nebula and well beyond the radio hot spot on the north-east side of the galaxy.}
\section{Discussion}
\subsection{Summary of results}
We summarise below some results derived in the previous section:
\begin{itemize}
   \item \emph {Extreme kinematics:} All the sources show disturbed off-nuclear emission line kinematics, including: line splitting in 3C 34 and 3C 330; and the detection of both broad (FWHM = 1000 -- 1500 km s$^{-1}$) and narrow (FWHM \ltae \ 600 km s$^{-1}$) components in the extended emission line regions of 3C 330, 3C 352 and 3C 435A. We emphasise the disturbed kinematics are detected in all the sources in our sample, not just those in which the radio and emission line structures have a similar scale.
   \item \emph {Relationship between linewidth and ionization state:} Although we do not see evidence for an anticorrelation between ionization state and linewidth as strong as seen in some low-redshift jet-cloud interaction sources (Clark et al. 1997, 1998; \pcite{villarmartin99b}), the low-ionization [OII]$\lambda$3727 and H$\beta$ emission lines tend to be broader than the high-ionization [OIII]$\lambda$5007 emission line in the regions where the complex kinematic structures described above are observed, especially in the central region of 3C 352. 
   \item \emph {Overall ionization state:} The ionization state peaks in the near-nucleus regions of the galaxies and decreases towards the EELR. In particular, in 3C 435A a minimum in the ionization state is observed at $\sim$ 2 arcsec north-east of the nucleus, just behind the radio hot spot.
   \item \emph {High electron temperatures:} The high electron temperatures measured in the extended regions of 3C 330 and in the central region of 3C 352 are inconsistent with photoionization by a central AGN \cite{tadhunter89a}, but they can be easily explained in terms of jet-cloud interactions \cite{villarmartin99b}.
   \item \emph {Extended UV continuum:} The contribution of the nebular continuum emission to the UV continuum is significant in all cases, with the nebular contribution ranging from $\sim$10 per cent in the nuclei of 3C 34 and 3C 435A, to $\sim$ 40 per cent in the extended regions of 3C 330. However, following subtraction of the nebular component, a significant UV excess remains in the extended nebulae in most of the objects.
\end{itemize}
\subsection{Kinematics: The evidence for shocks}
In previous sections we have presented evidence for extreme emission-line kinematics in all the objects. Below we discuss the implications of these results.

Line splitting is observed in the western EELR of 3C 34 and 3C 330 ($\Delta v \sim$ 1000 km s$^{-1}$ and $\sim$ 500 km s$^{-1}$, respectively). These velocity shifts are larger than what we would expect from gravity of an isolated elliptical galaxy ($\Delta v <$ 400 km s$^{-1}$) (e.g. \pcite{tadhunter89b}, \pcite{baum90}). Previously, high-velocity components have also been observed in the extended regions of powerful radio galaxies \cite{tadhunter91}, both at low redshifts (in Cygnus A) and at high redshifts (in 3C 265). Such splittings, which are characteristic of expanding shells of material, are most likely the result of interactions between the radio plasma and the interstellar medium. 3C 34 is a large double radio source, which extends 46 arcsec (365 kpc); however, the line splitting is detected at a radial distance of $\sim$ 2 to $\sim$ 4.5 arcsec (16 to 36 kpc) from the nucleus. Similarly, 3C 330 is a large 62 arcsec (458 kpc) double radio source, but we detect a line splitting in [OII] at a radial distance of $\sim$ 2.5 to $\sim$ 7 arcsec (18 to 51 kpc) from the nucleus. This indicates that the effects of the interactions between the radio components and the clouds of gas are also important in galaxies with extended radio sources, in which the main radio emitting components have passed beyond the EELR.
  
As mentioned above, these splittings are too large to be due to gravitational motions of the host galaxies. On the other hand, it is not yet clear whether direct shock acceleration associated with the radio components is sufficient to accelerate the clouds to such high velocities \cite{villarmartin99b}. Considering the warm clouds (T $\sim 10^{4}$ K) to be in equilibrium with a hot ambient phase (T $\sim 10^{7}$ K), the density contrast between the two phases is $\sim$ 1000, and the velocity of the clouds due to the advance of the radio jet is $v_{c} \sim v_{s} \sqrt{\frac{\rho_{h}}{\rho_{c} }}$ ($v_{s}$: shock velocity; $\rho_{h}$: density of the hot phase; $\rho_{c}$: density of the warm clouds). Assuming that the hot spot advance speed is $\sim$ 0.01$c$ --- 0.1$c$ \cite{scheuer95}, the clouds will be accelerated to velocities of $\sim$ 100 --- 1000 km s$^{-1}$ in the direction of the jet, but the velocities perpendicular to the radio axis will be much smaller. By measuring the ratio between the length of the radio arms and the extension perpendicular to the radio axes, the lateral velocities can be estimated. We obtain lateral extensions relative to the lengths of the radio arms of $\sim$ 0.4 in the case of 3C 34, and $\sim$ 0.15 for 3C 330 (see radio maps in \pcite{best97} and \pcite{mccarthy91}, respectively), implying lateral velocities of $\sim$15 --- 400  km s$^{-1}$. Given that the radio sources 3C 34 and 3C 330 are likely to be close to edge-on and on much larger scale than the emission-line structures, we expect the high-velocity components observed in these galaxies to be due to the lateral expansion of the cocoons. Therefore, the bowshock acceleration may not be sufficient to explain the large velocities. Some entrainment of the clouds in the post-shock wind (\pcite{klein94}; \pcite{villarmartin99b}) and/or in the turbulent boundary layers of the radio jets is required. Furthermore, fast winds from the quasars and starbursts associated with the host galaxies could also contribute to the large velocities observed in these sources \cite{heckman90}.
  
Further evidence for jet-cloud interactions is provided by the existence of broad kinematic components in the emission lines of some of the radio galaxies. We have detected underlying broad components with large linewidths (FWHM $\sim$ 1000 -- 1500 km s$^{-1}$) in the line profiles of three galaxies in our sample (3C 330, 3C 352 and 3C 435A), in addition to narrow components (FWHM \ltae \ 600 km s$^{-1}$). The large 62 arcsec (458 kpc) radio source 3C 330 shows an underlying broad component in both southern and northern EELR, at radial distances of $\sim$ 1.5 -- 4 arcsec (11 -- 30 kpc) from the nucleus. Again, evidence for interaction between the radio and optical structures can be detected in the EELR, even when the major radio components have passed beyond the EELR. In the case of 3C 352, in which the peaks of the radio lobes are located just outside the EELR, an underlying broad component has also been detected at both sides of the nuclear region. 3C 435A is the only object of our sample in which the radio hot spots have not passed yet beyond the EELR. In this object, a broad kinematic component can be detected in [OII]$\lambda3727$ at both sides of the nucleus. In all of these cases, the linewidths of the broad components are larger than would be expected for gravitational acceleration in the potential of a giant elliptical galaxy.

Similar kinematically disturbed components have also been observed at low redshifts in other powerful radio galaxies with jet-cloud interactions. High-dispersion long-slit spectra of the low-redshift radio galaxies 3C 171 \cite{clark98} and PKS 2250-41 \cite{villarmartin99b} have revealed multi-component emission-line kinematics: narrow components, with high-ionization state, and broad components, with low-ionization state. In the case of PKS 2250-41, the narrow component extends beyond the radio structure, while the broad component is found to be spatially associated with the radio emitting structures and is believed to represent gas cooling behind the shock fronts. As argued by \scite{villarmartin99b}, it is likely that the gaseous component associated with the broad lines is accelerated by entrainment in the faster moving hot ISM behind the shock front.

\begin{figure*}
\begin{minipage}{85mm}
\psfig{figure=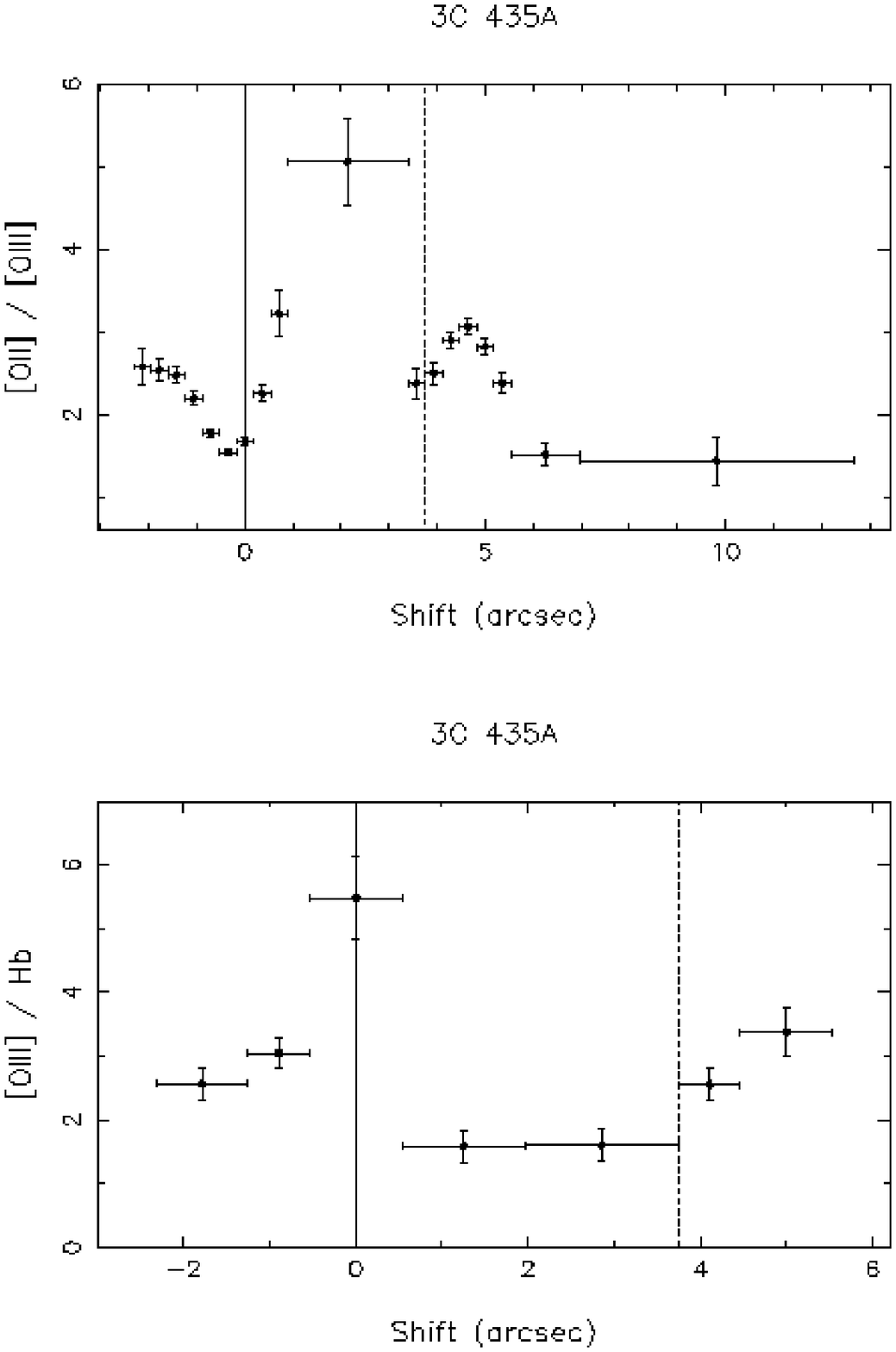,width=8.5cm}
\caption{Variation in the line ratios [OII]/[OIII] and [OIII]/H$\beta$ along the radio axis of 3C 435A.}
\label{435aratios}
\end{minipage}
\end{figure*}

However, we have to be cautious in interpreting the broad component solely in terms of jet-induced shocks, since a broad component is detected well beyond the northern radio hot spot in 3C 435A. This is similar to the phenomenon observed in the high-redshift radio galaxy MRC1558-003, which presents high velocities and large linewidths in the EELR beyond the radio structures \cite{villarmartin99a}. This suggests that another accelerating mechanism, apart from jet-cloud interactions, may be at work. Nonetheless, even in the case of 3C 435A, the \emph {broadest} emission lines are measured in a region behind the radio hot spot. 
Another common characteristic in the kinematics of the galaxies in our sample is that the [OII]$\lambda3727$ emission line tends to be broader than the [OIII]$\lambda5007$ emission line in the regions where the disturbed kinematics are observed. This could be because the [OII] line is a doublet and it was fitted with a single Gaussian. To quantify this broadening, we simulated [OII] emission lines of different widths and fitted them with both single and two Gaussians. We found that fitting the two components of the doublet with a single Gaussian made the line only $\sim$ 50 km s$^{-1}$ broader than if the two components were fitted separately. However, it happens that [OII] is broader by approximately 300 km s$^{-1}$ in 3C 34 at the location of the splitting, and in 3C 330 where the broad component is detected, corresponding also to low-ionization regions (see Figures~\ref{34ro2o3} and \ref{330ratios} (bottom), respectively). In the case of 3C 352, the [OII] line is observed to be $\sim$ 200 km s$^{-1}$ broader than [OIII] in the central region, within $\sim$ 1 arcsec distance on either side of the nucleus, corresponding with the region where the underlying broad component is detected. In 3C 435A, the [OII] emission line is observed to be much broader than [OIII] in the region behind the northern hot spot, coincident with a low-ionization state (see Figure~\ref{435aratios}). In 
addition, at a radial distance of $\sim$ 6 -- 12 arcsec north of the nucleus of 3C 435A, beyond the northern radio hot spot, and coincident with a high-ionization-state region (see Figure~\ref{435aratios} (top)), the [OII] line appears to be $\sim$ 500 km s$^{-1}$ broader than [OIII]. However, since the S/N ratio is low in this extended region, the linewidth of [OIII]$\lambda5007$ in this object is likely to be affected by the subtraction of the night-sky line that falls on the [OIII] emission line. 

An anticorrelation between linewidth and ionization state is a common feature of radio galaxies in which jet-cloud interactions are taking place: low-ionization lines are broader than high-ionization lines. This phenomenon has been observed in the powerful low-redshift radio galaxies 3C 171 \cite{clark98} and PKS 2250-41 (\pcite{clark97}, \pcite{villarmartin99b}). This anticorrelation is difficult to explain in terms of the AGN illuminating the undisturbed ambient gas in the host galaxy, but it can be easily explained in terms of shocks. The line emission contains contributions from both high-ionization-state gas, with narrow velocity dispersion, emitted in the undisturbed precursor zone, and low-ionization-state gas, with a much broader velocity distribution, emitted by the disturbed cooling gas behind the shock. Given the similarity between the kinematical properties of the objects in our sample and the low-redshift jet-cloud interaction objects, it is plausible that the broad and narrow components in our sample arise in a similar way at all redshifts.

Overall, the complex kinematics observed in the extended regions of the galaxies in our sample indicate that the material has been highly perturbed and accelerated by the interaction between the radio emitting components and the ambient gas. We see evidence for this shock acceleration even in those galaxies with the more extended radio sources (3C 34 and 3C 330), in which the radio hot spots have passed well beyond the EELR.
\subsection{Ionization mechanism: AGN photoionization or jet-induced shocks ?}
Although it is likely that a component of the ISM has been accelerated by jet-induced shocks in all the sources in our sample, this does not necessary imply that the clouds are shock-ionized. For example, clouds could be accelerated and compressed by jet-induced shocks, but then photoionized by an illuminating AGN in the core of the host galaxy after they have cooled. To address the issue of the ionization mechanism, in this section we compare the measured line ratios with the predictions of both AGN photoionization and shock-ionization models.

\begin{figure*}
\begin{minipage}{160mm}
\psfig{figure=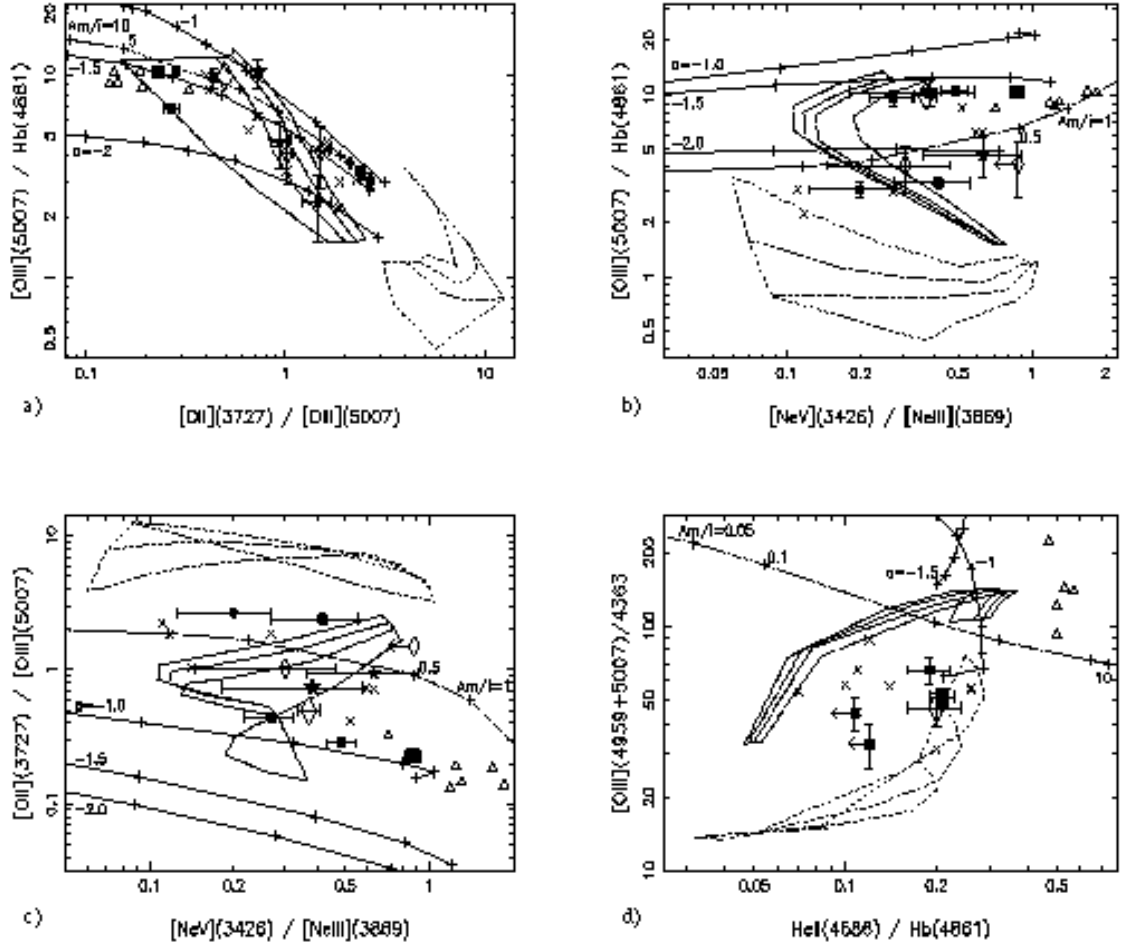,width=16cm}
\caption[]{Diagnostic diagrams showing the following line ratios: a)~[OIII](5007)/H$\beta$~vs~[OII](3727)/[OIII](5007); b) [OIII](5007)/H$\beta$ vs 
[NeV](3426)/[NeIII](3869); c) [OII](3727)/[OIII](5007) vs [NeV](3426)/[NeIII](3869); and d) [OIII](4959+5007)/4363 vs HeII(4686)/H$\beta$(4861) for the four 
sources: 3C 34 (stars), 3C 330 (filled squares), 3C 352 (open diamonds) and 3C 435A (filled circles). Bigger symbols denote the nuclear regions and smaller 
ones the extended regions of the galaxies. For comparison, line ratios corresponding to the EELR of low-redshift radio galaxies are also plotted; 
shock-ionized objects {\small (crosses)} taken from \scite{clark96}: PKS 2250-41, 3C 171, 4C 29.30 and Coma A, and from \scite{villarmartin98}: PKS 1932-464; 
and the photoionized galaxy 3C 321 {\small (open triangles)} taken from \scite{robinson2000}. Pluses linked by solid lines represent line ratios produced by 
optically-thick single-slab power-law ($F_{\nu} \propto \nu^{\alpha}$) photoionization models (using the code MAPPINGS) with spectral indices of $\alpha$ = 
-1.0, -1.5 and -2.0, and a sequence in the ionization parameter covering the range $5 \times 10^{-4} < U < 10^{-1}$. Pluses linked by a dot-dash-dot line 
indicate the predictions of the photoionization including matter-bounded clouds model from \scite{binette96}, with the ratio A$_{M/I}$ in the range 10$^{-4} 
\leq$ A$_{M/I} \leq$ 10. U and A$_{M/I}$ increase from right to left in a), and from left to right in b), c) and d). Predictions of pure shocks (dashed 
lines), and 50 per cent shocks + 50 per cent precursor models (dotted lines) from Dopita \& Sutherland (1995, 1996) are also plotted, each sequence 
corresponding to a fixed magnetic parameter (B/$\sqrt{n}$ = 0, 1, 2, 4 $\mu$Gcm$^{-3/2}$) and a changing shock velocity in the range 150 $\leq v_{s}\leq$ 500 
km s$^{-1}$).}
\label{diagmosaic}
\end{minipage}
\end{figure*}

Figure~\ref{diagmosaic} shows the resulting diagnostic diagrams, which involve the line pairs: [NeV]$\lambda$3426 \& [NeIII]$\lambda$3869, [OII]$\lambda$3727 \& [OIII]$\lambda$5007, H$\beta$ \& [OIII]$\lambda$5007, HeII(4686) \& H$\beta$ and the electron temperature diagnostic line ratio [OIII](4959+5007)/4363. Emission line pairs involving the same element are particularly useful for the diagnostic diagrams because the dependence of the line ratios on the elemental abundances is relatively minor \cite{baldwin81}. Also, if the two emission lines in the pair are close in wavelength, the possible errors arising from calibration and reddening corrections are minimized.  

To generate the AGN model predictions we used the multipurpose photoionization code MAPPINGS. The results are plotted as solid lines, which represent the line ratios produced by optically-thick, single-slab, power-law photoionization models ($F_{\nu} \propto \nu^{\alpha}$) with spectral indices of $\alpha$ = -1.0, -1.5 and -2.0 (from top to bottom), and a sequence in the ionization parameter covering the range $2.5 \times 10^{-4} < U < 10^{-1}$. A density of 100 cm$^{-3}$ and solar abundances were assumed. 

The shocks models are taken from Dopita \& Sutherland (1995, 1996). We include grids of simple shocks (dashed lines) and shocks with the photoionized precursor included (50 per cent shocks + 50 per cent precursor) (dotted lines). The two main parameters for controlling the post-shock emission line spectrum are the velocity of the shock (150 $\leq v_{s} \leq$ 500 km s$^{-1}$) and the magnetic parameter (0 $\leq$ B/$\sqrt{n}$ $\leq$ 4 $\mu$Gcm$^{-3/2}$). Each sequence in the Figures corresponds to a fixed B/$\sqrt{n}$ value and a changing $v_{s}$. \nocite{dopita95} \nocite{dopita96}

In addition to the photoionization and shock models, we have also plotted mixed-medium photoionization models, including both matter-bounded and radiation-bounded clouds (dot-dash-dot line) from \scite{binette96}. These combine an optically-thick zone (ionization-bounded [IB] component), illuminated by a photon spectrum that is initially filtered through an optically-thin zone (matter-bounded [MB] component). The sequences of line ratios were derived by allowing the ratio (A$_{M/I}$) of the solid-angle subtended from the nuclear photoionizing source by the MB clouds relative to that subtended by the IB clouds to vary in the range 10$^{-4} \leq$ A$_{M/I} \leq$ 10. Since by definition the MB clouds lie somewhere in between the central source and the IB clouds, strictly speaking, A$_{M/I}$ cannot be lower than unity. However, based on the unified schemes and due to orientation effects, some fraction of the MB clouds located in the inner regions could be obscured from the observer (e.g. by a dusty torus), therefore giving a physical meaning to an apparent A$_{M/I} <$ 1.

For comparison, we have also included data corresponding to the EELR of low-redshift radio galaxies. The crosses represent objects in which strong jet-cloud interactions are taking place: PKS 2250-41, 3C 171, 4C 29.30, Coma A \cite{clark96}, and PKS 1932-464 \cite{villarmartin98}. The open triangles represent the photoionized radio galaxy 3C 321 which is consistent with mixed-medium photoionization models \cite{robinson2000}.

Figure~\ref{diagmosaic} (a) shows the [OIII](5007)/H$\beta$ line ratio plotted against [OII](3727)/[OIII](5007). This combination of emission line pairs was first used by \scite{baldwin81} to distinguish between the four predominant ionization mechanisms (HII regions, planetary nebulae, power-law photoionization and shock-wave heating). This diagram, which clearly distinguishes between pure-shock and photoionization models, shows that none of the galaxies fall on the region occupied by the pure shock-ionization predictions (dashed lines). We can see that the power-law photoionization predictions with $\alpha$ = -1.5 provide good fit to the data, in agreement with the results that Robinson etal (1987) obtained for a sample of low-redshift radio galaxies. However, we cannot rule out the mixed-medium photoionization and shocks+precursor models because they overlap with the power-law photoionization predictions in the diagnostic diagram.

To separate these ionization mechanisms from each other, in Figure~\ref{diagmosaic} (b), we have plotted the [OIII](5007)/H$\beta$ line ratio against [NeV](3426)/[NeIII](3869) on a diagnostic diagram taken from \scite{dopita97}. Although this diagram appears to discriminate better between photoionization and shock+precursor models, it also fails to offer a clean separation. In this case, the data fall in the region occupied by the photoionization ($\alpha$ = -1.5 and -2.0), mixed-medium photoionization model, and shock+precursor predictions. 

Given that there is still some overlap between the models, we combined the [OII](3727)/[OIII](5007) line ratio and [NeV](3426)/[NeIII](3869) on a new diagnostic diagram shown in Figure~\ref{diagmosaic} (c). This diagram offers a much cleaner separation between photoionization and shock+precursor models. The galaxies fall mainly in the region occupied by the shock+precursor models (dotted lines). However, mixed-medium photoionization model cannot be ruled out because it overlaps with the shock+precursor predictions.

Figure~\ref{diagmosaic} (d) shows the electron temperature diagnostic line ratio [OIII](4959+5007)/4363 against HeII(4686)/H$\beta$ on a diagnostic diagram. Only for 3C 330 and 3C 352, was it possible to measure the line ratio [OIII](4959+5007)/4363, and by using this ratio, we derived the electron temperature for both sources, assuming a low-density limit. The temperatures obtained, 14900 $\leq$ T$_{e}$ $\leq$ 23000 K, are too high to be explained by pure photoionization models, which generally predict electron temperatures less than 11000 K \cite{wilson97}. Mixed-medium photoionization models can give high temperatures, but not for such low ionization \cite{binette96}. The same problem has been found in the extended regions of other radio galaxies \cite{tadhunter89a}. On the other hand, the temperatures are consistent with shock models which include some precursor. Note that we cannot rule out a contribution from high-density gas in the nuclear regions of 3C 330 and 3C 352 which might lead us to overestimate the temperature if we assume a low density limit. However, there is currently no strong observational evidence for a high-density inner narrow-line region in well resolved low-redshift radio galaxies (\pcite{tadhunter94}, \pcite{robinson2000}). 

By comparing these three diagnostic diagrams ((a), (b) and (c)), we see that pure-shock predictions are inconsistent with the data, and it appears that both mixed-medium photoionization and shock+precursor models provide a good fit. However, in the temperature diagnostic diagram (d) the data appear to be more consistent with the pure-shock than with any of the other models. 

The result obtained for 3C 352 (consistent with shock+precusor predictions), which is based on emission line diagnostics at rest-frame optical wavelengths, is in agreement with that obtained by \scite{best2000a}, who used a diagnostic diagram based on rest-frame UV line ratios ([CIII](2326)/[CII](1909) \& [NeIII](3869)/[NeV](3426)).
 
There is a general inconsistency from diagram to diagram for individual sources, as well as for the ensemble of points for all the galaxies together. This lack of consistency shows the danger in basing the conclusions on the results obtained from a single diagnostic diagram, and suggest that some of the assumptions implicit in the models may be wrong. If we consider the temperature diagnostic diagram, it can be seen that the [OIII](4959+5007)/4363 line ratios of the data are lower, and nearer to the pure-shock predictions than we would expect on the basis of the other diagrams. This discrepancy could be explained if we consider that the post-shock clouds are destroyed by the interaction with the hot post-shock wind as they cool \cite{klein94} -- the high-velocity components may be symptomatic of this destruction process. By reducing the size of the post-shock cooling zone, weaker emissions in [OII] and H$\beta$ would be measured, while the [OIII] emission would not be affected. Because of this, we expect to measure lower [OII]/[OIII] ratios and higher [OIII]/H$\beta$ ratios than predicted by the models, but we would expect the [OIII](4959+5007)/4363 ratio to be unaffected. This could explain the inconsistency between the results of the temperature diagnostic diagram (d), which involves the relatively high-ionization [OIII] and HeII emission lines and shows the data near the pure-shock predictions, and diagrams (a), (b) and (c), which also involve the [OII] and H$\beta$ emission lines. This idea of ``matter-bounded shocks'' needs to be checked with more detailed modelling.
\subsection{Comparison with low-redshift radio galaxies}
In this section we discuss the similarities and differences between the four sources studied in this paper and the low-redshift jet-cloud interaction radio galaxies.  

The four galaxies we have studied in this paper show many of the properties observed in the low-redshift jet-cloud interactions, including: a) complex emission-line kinematics ; b) low-ionization state in the EELR; c) high electron temperatures in the EELR; and d) a UV continuum excess.

However, the major difference is that the galaxies of our sample exhibit those properties in a more moderate way than the low-redshift objects. Perhaps, this is not surprising given that we are comparing the high-redshift galaxies with the most extreme jet-cloud interaction objects in the low redshift regime. 

In particular, a common feature in the sources of the low-redshift sample is an anticorrelation between linewidth and ionization state. However, in the galaxies of our sample we do not see such strong evidence for this anticorrelation. It is only in the central region of 3C 352 that we find convincing evidence for such anticorrelation. 

Low-redshift radio galaxies studied by \scite{clark96}, such as 3C 171, PKS 2250-41 and 4C 29.30, present a UV continuum excess that is mainly dominated by the nebular continuum emission. However, in the objects of our sample a very significant UV continuum excess still remains after the subtraction of the corresponding nebular contribution. Other possible explanations for this excess include jet-induced star formation, scattered light by dust or electrons emitted by the obscured AGN (although this latter explanation is unlikely at such large distances from the nuclear region), and merger-induced star formation \cite{tadhunter97}.

In Figure~\ref{diagmosaic} it can be seen that the low-redshift jet-cloud interaction objects (denoted by `$\times$') fall over the same region as the four galaxies of our sample. This indicates that the main ionization mechanism of the sources in both samples is likely to be the same. In contrast, the points corresponding to 3C 321 (denoted by `$\triangle$') fall on a well-defined region, clearly separated from the other galaxies and consistent with the mixed-medium photoionization models, showing that 3C 321 is mainly photoionized, with little or no contribution from shock ionization \cite{robinson2000}, unlike the rest of the sources presented in the diagnostic diagrams.
\section{Conclusions}
We report the results obtained from a study, based on long-slit spectroscopy, of the kinematics and ionization mechanisms of the line-emitting gas for a small sample of high/intermediate-redshift radio galaxies (3C 34, 3C 330, 3C 352 and 3C 435A). 

We see evidence for shock-acceleration of the emission-line gas in the EELR of all the galaxies, even in the largest radio sources of our sample (3C 34 and 3C 330), in which the radio hot spots have passed the extended gas of the galaxies. The extended regions present highly disturbed kinematics (line splitting and underlying broad components), which are difficult to explain if we do not consider a strong interaction between the radio-emitting components and the ambient gas. 

In contrast, we do not find consistency in explaining the dominant ionization mechanism of the line-emitting gas. We have compared the emission-line ratios with a set of theoretical predictions, including: power-law photoionization models, mixed-medium photoionization models, pure-shock models and shocks including a photoionized precursor. Power-law photoionization models and pure-shock models fail to fit the data well. We find that shock+precursor predictions provide the best fit overall. However, there is a lack of consistency from diagram to diagram for some of the sources. Particularly problematic is the diagnostic diagram involving the temperature diagnostic line ratio [OIII](4959+5007)/4363. This inconsistency between the different diagrams suggest that some of the assumptions of the models might be wrong.

Further spectroscopic investigation of these sources is required. Infrared spectroscopic analysis would allow us to measure important diagnostic emission lines, such as [OI]$\lambda6300$, H$\alpha$, [NII]$\lambda\lambda6548,6583$ and [SII]$\lambda\lambda6717,6731$. The corresponding line ratios would be compared with both AGN-photoionization and shock-ionization predictions, in order to further constrain the main ionization mechanism of the extended gas in these sources. Furthermore, deep UV imaging will allow us to understand the nature of the observed extended UV continuum, as well as to study the environment of these radio galaxies.
\section*{Acknowledgments}
This work is based on observations made at the Observatorio del Roque de los Muchachos, La Palma, Spain. We thank John Dyson for reading the manuscript and providing useful discussions. CSI acknowledges a White Rose studentship.

\bibliographystyle{mnras}
\bibliography{reference}

\begin{table}
\begin{tabular}{cccccccccc}\hline\hline \\
{\bf Dates }&{\bf Object}&{\bf t$_{exp}$}&{\bf PA}&{\bf Grat.}&{\bf $\lambda$ Range}&{\bf Pixel Scale}&{\bf Resolut.}&{\bf Slit}&{\bf Seeing} \\
   & &(s)& (deg.) & & (\AA) & (arcsec/\AA)& (\AA)&(arcsec)&(arcsec)\\ \\ \hline \\
15/8/93 & 3C34 & 3600 & 88 & 158R & 5691-9065 & 0.335/2.72 & 9.0$\pm$0.2 & 1.5 & 1.2 \\ 
20/7/96 & 3C330 & 6000 & 230 & 158R & 6247-9616 & 0.335/2.73 & 7.4$\pm$0.3 & 1.35 & 1.0 \\ 
20/7/96 & 3C330 & 6000 & 230 & 300B & 4853-6316 & 0.357/1.54 & 4.28$\pm$0.04&1.35 & 1.0 \\ 
21/7/96 & 3C352 & 6000 & 140 & 158R & 7175-9823 & 0.335/2.73 & 10.1$\pm$0.3 & 1.56 & 1.3 \\ 
21/7/96 & 3C352 & 6000 & 140 & 158B & 4599-7549 & 0.357/2.89 &10.31$\pm$0.06&1.56 & 1.3 \\ 
21-22/7/96&3C435A&9900 & 229 & 158R & 6455-9825 & 0.335/2.73 & 8.9$\pm$0.5 &1.56 & 1.2 \\ 
22/7/96 &3C435A & 6000 & 229 & 158B & 3753-6550 & 0.357/2.89 & 9.63$\pm$0.09&1.56 & 1.0 \\ \\ \hline\hline
\end{tabular}
\caption{Log of the spectroscopic observations.}
\label{logobs}
\end{table}

\begin{table}
\begin{tabular}{ccc}\hline\hline \\
{\bf Object }&{\bf Redshift}&{\bf Angular scale} \\
 & & (kpc arcsec$^{-1}$ ) \\ \\ \hline \\
3C 34 & 0.690 & 7.94\\
3C 330 & 0.550 & 7.39\\
3C 352 & 0.8067 & 8.24\\
3C 435A & 0.471 & 6.94\\ \\ \hline\hline
\end{tabular}
\caption{Angular scales.}
\label{angscale}
\end{table}

\begin{table}
\begin{tabular}{lccccc}\hline\hline \\
{\bf Wavelength (\AA)}&{\bf Line}& &{\bf Eastern}&{\bf Nucleus}&{\bf Western}\\ 
 & & &{\bf EELR}& &{\bf EELR}\\ \\ \hline \\
3426 \dotfill& [NeV] & & --- & 0.32$\pm$0.17 & 0.54$\pm$0.24 \\
3727 \dotfill& [OII] & & 8.59$\pm$7.84 & 7.54$\pm$1.21 & 4.36$\pm$1.07 \\
3869 \dotfill& [NeIII] & & --- & 0.84$\pm$0.16 & 0.83$\pm$0.26 \\
3967 \dotfill& [NeIII] & & --- & 0.38$\pm$0.10 & --- \\
4340 \dotfill& H$\gamma$ & & --- & 0.53$\pm$0.15 & --- \\
4861 \dotfill& H$\beta$ & & 1.00$\pm$0.92 & 1.00$\pm$0.16 & 1.00$\pm$0.24 \\
4959 \dotfill& [OIII] & & 2.39$\pm$2.19 & 3.44$\pm$0.55 & 1.55$\pm$0.39 \\
5007 \dotfill& [OIII] & & 7.17$\pm$6.57 & 10.31$\pm$1.65 & 4.66$\pm$1.18 \\
5199 \dotfill& [NI] & & --- & 0.69$\pm$0.26 & --- \\ \\ \hline \\
H$\beta$ flux (10$^{-16}$ erg cm$^{-2}$ s$^{-1}$) &\dotfill& & 0.39$\pm$0.36 & 1.47$\pm$0.24 & 0.80$\pm$0.19 \\ \\ \hline \\
{\bf Nebular Emission} & & & $\sim$ 11 \% & $\sim$ 12 \% & --- \\ \\ \hline\hline \\
\end{tabular}
\caption{Emission-line fluxes for 3C 34, including the eastern EELR (8.4 x 1.5 arcsec$^{2}$ aperture centred at 6 arcsec east of the nucleus), the nucleus (4.7 x 1.5 
arcsec$^{2}$ aperture centred at 0.5 arcsec west of the continuum centroid) and the western EELR (3.0 x 1.5 arcsec$^{2}$ aperture centred at 4.3 arcsec west 
of the nucleus). The nebular emission contribution to the continuum, for the wavelength range 3400 \AA \ -- 3700 \AA, is also given.}
\label{tabflux34}
\end{table}

\begin{table}
\begin{tabular}{lccccccc}\hline\hline \\
{\bf Wavelength (\AA)}&{\bf Line}& & {\bf Southern}&{\bf Nucleus}&{\bf Nucleus}&{\bf Northern}&{\bf Northern}\\ 
 & & & {\bf EELR}& & Red. corr. &{\bf EELR (n1)} &{\bf EELR (n2)}\\ \\ \hline \\
3346 \dotfill& [NeV]&  & --- & 0.113$\pm$0.015 & 0.24$\pm$0.03 & 0.15$\pm$0.03 & --- \\
3426 \dotfill& [NeV] & & --- & 0.37$\pm$0.03 & 0.73$\pm$0.06 & 0.36$\pm$0.04 & 0.32$\pm$0.07 \\
3727 \dotfill& [OII]&  & 1.84$\pm$0.41 & 1.43$\pm$0.08 & 2.37$\pm$0.13 & 2.98$\pm$0.14 & 4.19$\pm$0.45 \\
3869 \dotfill& [NeIII]& & --- & 0.54$\pm$0.04 & 0.84$\pm$0.06 & 0.74$\pm$0.06 & 1.19$\pm$0.17 \\
3967 \dotfill& [NeIII]&  & --- & 0.12$\pm$0.02 & 0.18$\pm$0.03 & 0.28$\pm$0.04 & 0.51$\pm$0.11 \\
4102 \dotfill& H$\delta$&  & --- & 0.135$\pm$0.018 & 0.19$\pm$0.02 & 0.27$\pm$0.03 & --- \\
4340 \dotfill& H$\gamma$&  & 0.33$\pm$0.12 & 0.36$\pm$0.02 & 0.46$\pm$0.03 & 0.50$\pm$0.04 & 0.50$\pm$0.09 \\
4363 \dotfill& [OIII]&  & 0.28$\pm$0.08 & 0.212$\pm$0.018 & 0.27$\pm$0.02 & 0.21$\pm$0.03 & 0.29$\pm$0.05 \\
4686 \dotfill& HeII&  & --- & 0.194$\pm$0.018 & 0.209$\pm$0.019 & 0.19$\pm$0.03 & --- \\
4861 \dotfill& H$\beta$&  & 1.00$\pm$0.20 & 1.00$\pm$0.06 & 1.00$\pm$0.06 & 1.00$\pm$0.04 & 1.00$\pm$0.10 \\
4959 \dotfill& [OIII]&  & 2.28$\pm$0.46 & 3.53$\pm$0.20 & 3.37$\pm$0.19 & 3.27$\pm$0.15 & 3.13$\pm$0.34 \\
5007 \dotfill& [OIII]&  & 6.83$\pm$1.38 & 11.11$\pm$0.61 & 10.37$\pm$0.57 & 10.50$\pm$0.47 & 9.62$\pm$1.02 \\
5721 \dotfill& [FeVII]&  & --- &0.19$\pm$0.04 & 0.13$\pm$0.03 & --- & --- \\ \\ \hline \\
H$\beta$ flux & & & & & & & \\
(10$^{-16}$ erg cm$^{-2}$ s$^{-1}$)&\dotfill&  & 1.13$\pm$0.23 & 5.60$\pm$0.31 &
--- & 1.79$\pm$0.08 & 1.27$\pm$0.13 \\ \\ \hline \\
{\bf Nebular Emission} & & & $\sim$ 39 \% & $\sim$ 50 \% & $\sim$ 25 \% & $\sim$ 37 \% & $\sim$ 32 \% \\ \\ \hline\hline \\
\end{tabular}
\caption{Emission-line fluxes for 3C 330, including the southern EELR (6.4 x 1.35 arcsec$^{2}$ aperture centred at 4.8 arcsec south-west of the continuum centroid), the nucleus 
(2.5 x 1.35 arcsec$^{2}$ aperture centred at 0.4 arcsec south-west of the continuum centroid), the northern EELR 'n1' (1.4 x 1.35 arcsec$^{2}$ aperture 
centred at 1.6 arcsec north-east of the nucleus) and the northern EELR 'n2' (2.5 x 1.35 arcsec$^{2}$ aperture centred at 3.5 arcsec north-east of the nucleus). Fluxes corresponding to the nucleus have been corrected for reddening (the errors do not take into account the uncertainity given by E$_{B-V}$). 
The nebular emission contribution to the continuum, for the wavelength range 3200 \AA \ -- 3700 \AA, is also given.}
\label{tabflux330}
\end{table}

\begin{table}
\begin{tabular}{lccccc}\hline\hline \\
{\bf Wavelength (\AA)}&{\bf Line}& &{\bf Southern}&{\bf Nucleus}&{\bf Northern}\\ 
 & & &{\bf EELR}& &{\bf EELR}\\ \\ \hline \\
2798 \dotfill& MgII&  & 0.15$\pm$0.06 & 0.38$\pm$0.05 & 0.58$\pm$0.23 \\
3346 \dotfill& [NeV] & & --- & 0.090$\pm$0.018 & --- \\
3426 \dotfill& [NeV]&  & 0.18$\pm$0.10 & 0.27$\pm$0.03 & --- \\
3727 \dotfill& [OII]&  & 4.18$\pm$1.19 & 4.89$\pm$0.47 & 6.25$\pm$2.04 \\
3869 \dotfill& [NeIII]&  & 0.59$\pm$0.22 & 0.73$\pm$0.08 & 0.94$\pm$0.35 \\
3886 \dotfill& H8+HeI&  & --- & 0.17$\pm$0.04 & --- \\
3967 \dotfill& [NeIII]&  & --- & 0.28$\pm$0.05 & --- \\
4102 \dotfill& H$\delta$&  & --- & 0.15$\pm$0.05 & --- \\
4340 \dotfill& H$\gamma$&  & 0.62$\pm$0.29 & 0.54$\pm$0.08 & --- \\
4363 \dotfill& [OIII]&  & --- & 0.28$\pm$0.05 & --- \\
4686 \dotfill& HeII&  & --- & 0.20$\pm$0.04 & --- \\
4861 \dotfill& H$\beta$&  & 1.00$\pm$0.29 & 1.00$\pm$0.10 & 1.00$\pm$0.32 \\
4959 \dotfill& [OIII]&  & 1.36$\pm$0.40 & 3.28$\pm$0.32 & 1.39$\pm$0.47 \\
5007 \dotfill& [OIII]&  & 4.09$\pm$1.19 & 9.83$\pm$0.95 & 4.17$\pm$1.42 \\ \\ \hline \\
H$\beta$ flux (10$^{-16}$ erg cm$^{-2}$ s$^{-1}$)&\dotfill&  & 1.45$\pm$0.42 & 3.60$\pm$0.35 & 0.87$\pm$0.28 \\ \\ \hline \\
{\bf Nebular Emission} & & & $\sim$ 37 \% & $\sim$ 37 \% & $\sim$ 28 \% \\ \\ \hline\hline \\
\end{tabular}
\caption{Emission-line fluxes for 3C 352, including the southern EELR (4.6 x 1.56 arcsec$^{2}$ aperture centred at 3.8 arcsec south-east of the continuum centroid), the nucleus 
(2.5 x 1.56 arcsec$^{2}$ aperture) and the northern EELR (3.2 x 1.56 arcsec$^{2}$ aperture centred at 2.8 arcsec north-west of the nucleus). The 
nebular emission contribution to the continuum, for the wavelength range 2600 \AA \ -- 3700 \AA, is also given.}
\label{tabflux352}
\end{table}

\begin{table}
\begin{tabular}{lcccccc}\hline\hline \\
{\bf Wavelength (\AA)}&{\bf Line}& &{\bf Southern}&{\bf Nucleus}&{\bf Northern}&{\bf Northern}\\
 & & &{\bf EELR}& &{\bf Component}&{\bf EELR}\\ \\ \hline \\
2798 \dotfill& MgII&  & --- & 0.36$\pm$0.08 & 0.44$\pm$0.10 & --- \\
3426 \dotfill& [NeV]&  & --- & 0.23$\pm$0.06 & 0.17$\pm$0.06 & --- \\
3727 \dotfill& [OII]&  & 8.50$\pm$4.96 & 7.86$\pm$0.66 & 8.04$\pm$0.74 & 3.42$\pm$1.22 \\
3869 \dotfill& [NeIII]&  & --- & 0.56$\pm$0.12 & 0.84$\pm$0.12 & --- \\
3967 \dotfill& [NeIII] & & --- & --- & 0.24$\pm$0.10 & --- \\
4861 \dotfill& H$\beta$&  & 1.00$\pm$0.54 & 1.00$\pm$0.08 & 1.00$\pm$0.09 & 1.00$\pm$0.34 \\ 
4959 \dotfill& [OIII]&  & --- & 1.10$\pm$0.09 & 1.00$\pm$0.10 & 0.79$\pm$0.29 \\
5007 \dotfill& [OIII]&  & --- & 3.29$\pm$0.28 & 3.01$\pm$0.29 & 2.36$\pm$0.86 \\
6300 \dotfill& [OI]&  & --- & 1.86$\pm$0.23 & 0.99$\pm$0.37 & --- \\
6548 \dotfill& [NII]&  & --- & 1.35$\pm$0.41 & 1.46$\pm$0.51 & --- \\
6563 \dotfill& H$\alpha$&  & --- & 4.22$\pm$0.57 & 3.70$\pm$0.66 & --- \\
6583 \dotfill& [NII]&  & --- & 4.35$\pm$0.56 & 3.90$\pm$0.63 & --- \\ \\ \hline \\
H$\beta$ flux (10$^{-16}$ erg cm$^{-2}$ s$^{-1}$)&\dotfill&  & 0.22$\pm$0.12 & 1.74$\pm$0.14 & 1.20$\pm$0.11 & 0.35$\pm$0.12 \\ \\ \hline \\ 
{\bf Nebular Emission} & & & --- & $\sim$ 10 \% & $\sim$ 21 \% & $\sim$ 17 \% \\ \\ \hline\hline \\
\end{tabular}
\caption{Emission-line fluxes for 3C 435A, including the southern EELR (12.1 x 1.56 arcsec$^{2}$ aperture centred at 9.1 arcsec south-west of the continuum centroid), the 
nucleus (5.4 x 1.56 arcsec$^{2}$ aperture), the northern component (4.6 x 1.56 arcsec$^{2}$ aperture centred at 4.6 arcsec north-east of the nucleus) and the 
northern EELR (5.7 x 1.56 arcsec$^{2}$ aperture centred at 9.8 arcsec north-east of the nucleus). The nebular emission contribution to the continuum, for the 
wavelength range 2600 \AA \ -- 3700 \AA, is also given.}
\label{tabflux435a}
\end{table}

\end{document}